\documentclass[10pt]{iopart}

\newcommand{\ket}[1]{| #1 \rangle}
\newcommand{\bra}[1]{\langle #1 |}

\usepackage{graphicx}
\usepackage{amsfonts}
\usepackage{amssymb}
\usepackage{bm}
\usepackage{latexsym}

\newtheorem{Def}{Definition}
\newtheorem{thm}[Def]{Theorem}
\newtheorem{lem}[Def]{Lemma}
\newtheorem{cor}[Def]{Corollary}

\newenvironment{prf}{\vspace{10pt} \noindent \textbf{Proof.} \ }
 {\hfill $\Box$ \par \vspace{10pt}}
\newenvironment{prf2}{\vspace{10pt} \noindent}
 {\hfill $\Box$ \par \vspace{10pt}}

\renewcommand{\theenumi}{\roman{enumi}}

\begin{document}

\title{Error exponents for entanglement concentration}

\author{Masahito Hayashi\dag, Masato Koashi\ddag, Keiji Matsumoto\S,
 Fumiaki Morikoshi$\|$ and Andreas Winter\P}
        
\address{\dag\ Laboratory for Mathematical Neuroscience, Brain Science
 Institute, RIKEN,\\ 2-1Hirosawa, Wako, Saitama, 351-0198, Japan}

\address{\ddag\ CREST Research Team for Interacting Carrier Electronics,
 School of Advanced Sciences, The Graduate University for Advanced Studies
  (SOKENDAI),\\ Hayama, Kanagawa, 240--0193, Japan}

\address{\S\ Quantum Computation and Information Project, ERATO, JST,\\ 5-28-3,
 Hongo, Bunkyo-ku, Tokyo,113-0033, Japan}

\address{$\|$\ NTT Basic Research Laboratories, NTT Corporation,\\ 3-1
  Morinosato-Wakamiya, Atsugi-shi, Kanagawa, 243-0198, Japan}

\address{\P\ Department of Computer Science, University of Bristol,\\
  Merchant Venturers Building, Woodland Road, Bristol BS8 1UB, United Kingdom}

\begin{abstract}

Consider entanglement concentration schemes that convert $n$ identical copies
of a pure state into a maximally entangled state of a desired size with success
probability being close to one in the asymptotic limit.
We give the distillable entanglement, the number of Bell pairs distilled per
copy, as a function of an \textit{error exponent}, which represents the rate of
decrease in failure probability as $n$ tends to infinity.
The formula fills the gap between the least upper bound of distillable
entanglement in probabilistic concentration, which is the well-known entropy
of entanglement, and the maximum attained in deterministic concentration.
The method of types in information theory enables the detailed analysis
of the distillable entanglement in terms of the error rate.
In addition to the probabilistic argument, we consider another type of
entanglement concentration scheme, where the initial state is
deterministically transformed into a (possibly mixed) final state whose
fidelity to a maximally entangled state of a desired size converges to one in
the asymptotic limit.
We show that the same formula as in the probabilistic argument is valid for
the argument on fidelity by replacing the success probability with the
fidelity.
Furthermore, we also discuss entanglement yield when optimal success
probability or optimal fidelity converges to zero in the asymptotic limit
(strong converse), and give the explicit formulae for those cases.

\end{abstract}

\pacs{03.67.-a, 03.67.Hk, 03.65.Ud}

\maketitle

\section{\label{Sec:Intro}Introduction}

Quantum entanglement, an indispensable resource for quantum information
processing such as superdense coding \cite{Bennett92}, quantum teleportation
\cite{Bennett93}, quantum cryptography \cite{Ekert}, and quantum computation
\cite{Jozsa}, is expected to have a rich mathematical structure behind its
weirdness.
As in the case of other physical resources, quantification of entanglement is
the key to understanding its full potential.
The essentials of bipartite pure-state entanglement have already been revealed
for both finite regimes and the asymptotic limit.
The fundamental results are the intimate connection between the mathematical
theory of majorization and entanglement manipulation
\cite{Nielsen99,Vidal99,Hardy99,Jonathan99}, and the existence of a unique
measure of entanglement in the asymptotic limit \cite{Bennett96,Popescu}.

One way of quantifying entanglement is to estimate the number of Bell pairs,
\begin{equation}
 \frac{1}{\sqrt{2}} (\ket{00}_{\rm AB} + \ket{11}_{\rm AB}),
\end{equation}
distilled from a given entangled state by local operations and classical
communication (LOCC).
Though the above quantity of distillable entanglement can be defined for mixed
states, we deal with only pure states here.
In order to make use of partially entangled states for quantum teleportation,
we need to convert the partially entangled states into maximally entangled
states by LOCC.
The process is called entanglement concentration, and its efficiency in the
asymptotic limit is the focus of this paper.

The unique measure of bipartite pure-state entanglement gives the limitation
on the efficiency of entanglement concentration.
Suppose we share $n$ identical copies of a partially entangled state
\begin{equation}
 \ket{\phi} = \sum_{i=1}^{d} \sqrt{p_i} \ket{i} \ket{i},
 \label{initial}
\end{equation}
where the Schmidt coefficients squared are arranged in decreasing order, i.e.,
$p_1 \geq p_2 \geq \cdots \geq p_d \geq 0$, and sum to one.
(Schmidt coefficients are arranged in decreasing order throughout this paper.)
Bennett \textit{et al.} \cite{Bennett96} proved that the maximum number of
Bell pairs distilled per copy from $\ket{\phi}^{\otimes n}$ is given by
\begin{equation}
 E_{\rm entropy}(\phi) = -\sum_{i=1}^d p_i \log_2 p_i
 \label{entropy}
\end{equation}
in the asymptotic limit, $n \to \infty$.
(Logarithms are taken to base two throughout this paper unless stated
otherwise.)
They imposed the condition that the success probability of entanglement
concentration tends to one in the asymptotic limit, i.e.,
\begin{equation}
 p_{\rm success} = 1- \epsilon,
\end{equation} 
where
\begin{equation}
 \epsilon \to 0  \qquad {\rm as} \qquad n \to \infty.
\end{equation}
With this restriction, the maximum attainable entanglement yield is proven
to be equation~(\ref{entropy}).

On the other hand, much research on entanglement concentration has been
undertaken from various viewpoints
\cite{Vidal99,Hardy99,Jonathan99,Lo,Morikoshi2000,Morikoshi2001,Hayashi2001,
Vidal2000}.
Among other things, the bound on entanglement yield in deterministic
concentration \cite{Morikoshi2001}
\begin{equation}
 E_{\rm det} (\phi)= - \log p_1
 \label{deterministic}
\end{equation}
gives another quantification of entanglement.
The restriction \textit{deterministic} means that the process succeeds with
probability one both in finite regimes and in the asymptotic limit.

Though the quantities $E_{\rm entropy}$ and $E_{\rm det}$ give
entanglement yield in the asymptotic limit, where both processes succeed with
probability one, the two quantities do not coincide.
The main purpose of this paper is to find out the reason for the discrepancy.
We will see that it is caused by the difference of the rate at which failure
probabilities decrease when $n$ tends to infinity in both concentration
processes.
Roughly speaking, while we obtain $E_{\rm entropy}$ when failure probability
decreases slowly, we obtain $E_{\rm det}$ when it decreases rapidly.
We will represent the rate by the exponent of failure probability in the
asymptotic limit (\textit{error exponent}).
This is a common approach in the information sciences, and will allow us to
`tune' between the two extremes just mentioned.
The quantum fixed-length pure state source coding in reference~
\cite{Hayashi2002} is another example that uses the notion of error exponents
in quantum information theory.

In the derivation of $E_{\rm entropy}$, we use the asymptotic equipartition
property \cite{CT}.
However, a detailed analysis of the asymptotic behaviour requires more powerful
mathematical tools; namely, the method of types \cite{CT,CK}, which makes it
possible to calculate the probabilities of rare events and derive stronger
results than when we focus only on typical sequences.

The argument via the method of types will give entanglement yield as
a function of an error exponent and reveal the missing link between
$E_{\rm entropy}$ and $E_{\rm det}$.
In addition, we will also see that the success probability exponentially
decreases when we try to distil more entanglement than $E_{\rm entropy}$
(strong converse).
This was observed in reference~\cite{Lo}, but here we are able to derive the
\textit{exact} error rate.

It was suggested that the two extremes $E_{\rm entropy}$ and
$E_{\rm det}$ can also be expressed as some limits of R\'{e}nyi entropy
\cite{Morikoshi2001}.
We will also see that they can be linked by R\'{e}nyi-entropy-like functions,
which is useful for practical calculation.

We can consider two different ways of analyzing entanglement concentration:
One is estimating the optimal success probability of obtaining the exact copy
of a maximally entangled state of a desired size, where we discard failure
cases.
The other is estimating the fidelity of the final state, which is generally
a mixed state, to a maximally entangled state of a desired size, where we
obtain the final mixed state with probability one.
We will first derive the entanglement yield as a function of an exponent of
failure probability (error exponent) under the condition that the optimal
success probability converges to one in the asymptotic limit.
Then, we will also derive the entanglement yield as a function of an exponent
of one minus fidelity, under the condition that the optimal fidelity converges
to one in the asymptotic limit.
The argument on fidelity can be reduced to that on probability via some
lemmata.
Finally, the yield functions will turn out to be in the same form in the both
cases.
Since transformations assumed in the argument on fidelity need not produce the
exact copy of a maximally entangled state, the treatment is more natural from a
physical perspective than that on optimal success probability.
In finite regimes, deterministically transforming a pure state into a maximally
entangled state with optimal fidelity has been discussed in
reference~\cite{Vidal2000}, while we will consider that in the asymptotic
limit.
The strong converse will also be analyzed in terms of fidelity.

This paper is organized as follows.
We begin in section \ref{Sec:Revisited} by revisiting the entanglement
concentration in finite dimensions.
After a brief review of the method of types in section~\ref{Sec:Types},
we will move on to the main result of this paper, error exponents and
asymptotic entanglement concentration, in section~\ref{Sec:Asymptotic}.
Section \ref{Sec:Strong} discusses the strong converse.
In section~\ref{Sec:Alternative formulae}, we will present alternative formulae
for the result proven in the preceding sections.
Finally, we will discuss interpretations of our result and of some properties
of the yield function, then conclude the paper.
Some lemmata used in the proofs of our results are presented in Appendices.

\section{\label{Sec:Revisited}Finite-dimensional entanglement concentration
 revisited}

In this section, we revisit entanglement concentration of finite-dimensional
states so that the asymptotic limit will be smoothly derived from it.
Suppose we distil a maximally entangled state with Schmidt number $L(\leq d)$,
\begin{equation}
 \ket{\Phi_L} = \frac{1}{\sqrt{L}} \sum_{i=1}^{L} \ket{i} \ket{i},
 \label{maximally}
\end{equation}
from the partially entangled state with Schmidt number $d$,
equation~(\ref{initial}).
Lo and Popescu \cite{Lo} derived the optimal probability with which we distil
$\ket{\Phi_L}$ from $\ket{\phi}$:
\begin{equation}
 P_L = \min_{l \in [1,L]} \frac{L}{L-l+1} \sum_{i=l}^d p_i.
 \label{Lo}
\end{equation}
Note that the Schmidt number of the initial state cannot be expanded by LOCC.
Thus, we cannot distil a maximally entangled state with the Schmidt number
greater than $d$.

In the following, we reformulate the optimal probability $P_L$ in a more
suitable form for our treatment of asymptotic entanglement concentration.
Our strategy for distilling a maximally entangled state consists of two parts:
\begin{enumerate}
 \item We perform a two-valued local measurement to change the initial
  distribution of the Schmidt coefficients (probabilistic part).
 \item If a desired result is obtained in the above measurement, we distil a
  maximally entangled state from the resultant state with probability one
  (deterministic part).
\end{enumerate}

First, we briefly review the second (deterministic) part, which was
investigated in reference~\cite{Morikoshi2001}.
Suppose we wish to distil a maximally entangled state with the greatest
possible Schmidt number from $\ket{\phi}$ with probability one.
The maximum Schmidt number of the maximally entangled state is
$\left\lfloor 1 / p_{1} \right\rfloor$, where $\lfloor x \rfloor$
represents the largest integer equal to or less than $x$.
Thus, if we could make $p_1$ smaller somehow, the size (Schmidt number) of the
resultant maximally entangled state would become greater.
Note that, according to Nielsen's theorem \cite{Nielsen99}, the largest Schmidt
coefficient cannot be deterministically decreased by LOCC.

In order to distil a maximally entangled state of size
$L > \left\lfloor 1 / p_{1} \right\rfloor$, we need to
adjust the largest Schmidt coefficient of the initial state before moving on to
deterministic entanglement concentration in the second part.
So, in the first (probabilistic) part, we perform the measurement presented
below to truncate the initial distribution of the Schmidt coefficients.

\begin{figure}[ht]
 \begin{center}
  \includegraphics{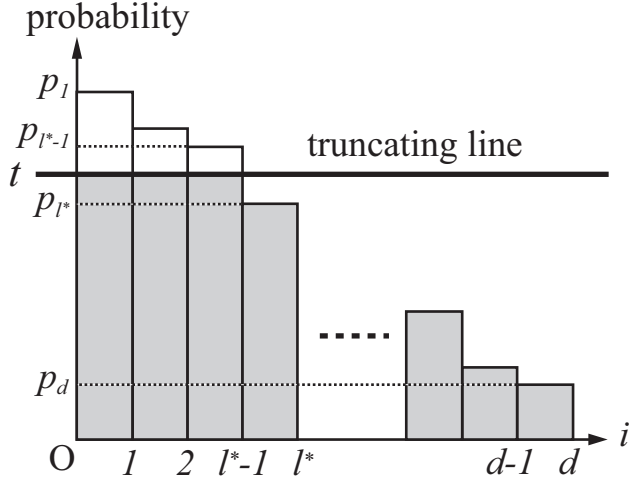}
 \end{center}
 \caption{\label{Fig1} The distribution of the squared Schmidt coefficients of
  a partially entangled state to be concentrated.
  Each bar corresponds to a Schmidt coefficient squared.
  The area of the shaded region below the truncating line $t$ represents the
  success probability of entanglement concentration.}
\end{figure}

As shown in figure~\ref{Fig1}, first we draw a truncating line that represents
probability $t$, which is uniquely determined by the size of the maximally
entangled state as proved later.
Then we perform a two-valued local measurement on either side of the entangled
pair by using measurement operators,
\begin{equation}
 M_1 = \sum_{i=1}^{l^*-1} \sqrt{\frac{t}{p_i}} \ket{i} \langle i | +
 \sum_{i=l^*}^{d} \ket{i} \langle i |,
 \label{measurement}
\end{equation}
with $p_1 \geq p_2 \geq \cdots \geq p_{l^*-1} > t \geq p_{l^*} \geq \cdots
\geq p_d$,
and $M_2$ such that $M_1^\dagger M_1 + M_2^\dagger M_2 = I$.
Measurement outcome 1 corresponds to the success of the entanglement
concentration.
It is easily seen that it occurs with probability
\begin{equation}
 P_{\rm success} = \parallel (M_1 \otimes I) \ket{\phi} \parallel^2 =
 \sum_{i=1}^{d} \min \{ t, p_i \},
 \label{truncation}
\end{equation}
which is shown schematically as the area of the shaded region below the
truncating line $t$ in figure~\ref{Fig1}.
Measurement outcome 2 corresponds to the failure of the
concentration, and the failure probability, $1-P_{\rm success}$, is equal
to the area of the region above the truncating line in figure~\ref{Fig1}.

In the case of measurement outcome 1, the post-measurement state becomes
\begin{equation}
 \ket{\phi'} = \frac{(M_1 \otimes I) \ket{\phi}}{\sqrt{P_{\rm success}}} =
 \frac{1}{\sqrt{P_{\rm success}}} \left( \sum_{i=1}^{l^*-1} \sqrt{t}
 \ket{i} \ket{i} + \sum_{i=l^*}^{d} \sqrt{p_i} \ket{i} \ket{i} \right),
 \label{postmeasurementstate}
\end{equation}
whose Schmidt coefficients squared are represented as the shaded region in
figure~\ref{Fig1} with an appropriate renormalization;
Each bar (Schmidt coefficient) is replaced with that divided by
$P_{\rm success}$.
Since the largest Schmidt coefficient of the post-measurement state is
$\sqrt{t/P_{\rm success}}$, we can distil a maximally entangled state of
size
\begin{equation}
 L = \frac{P_{\rm success}}{t}.
 \label{LvsP}
\end{equation}
In other words, our scheme distils a maximally entangled state of size $L$
with probability $P_{\rm success}$.
The truncating probability $t$ is uniquely determined by a
given integer $L$ via an implicit function
\begin{equation}
 L = \sum_{i=1}^{d} \min \left\{ 1, \frac{p_i}{t} \right\},
 \label{Lvst}
\end{equation}
which is given by equations~(\ref{truncation}) and (\ref{LvsP}).
Since the right-hand side of equation~(\ref{Lvst}) is strictly monotone
decreasing in $\left[1 / p_{1}, d \right]$ for $t \in [p_d, p_1]$, $t$ is
uniquely determined for a given integer $L \in \left[1 / p_{1}, d \right]$.
(When $L< 1 / p_{1}$, $\ket{\Phi_L}$ can be distilled with probability one.
Thus, we do not need the probabilistic part.)

The rest of this section gives the proof that the probability
$P_{\rm success}$ coincides with the optimal probability $P_L$
[equation~(\ref{Lo})].
The truncating probability $t$ uniquely determines an integer $l^*$ satisfying
$p_{l^*-1} > t \geq p_{l^*}$. (See figure~\ref{Fig1}.)
From equations~(\ref{truncation}) and (\ref{LvsP}), we have
\begin{equation}
 P_{\rm success} = t(l^*-1) + \sum_{i=l^*}^d p_i = tL.
 \label{determining-t}
\end{equation}
Thus,
\begin{equation}
 t=\frac{1}{L-l^*+1} \sum_{i=l^*}^d p_i.
 \label{t}
\end{equation}
Substituting this into equation~(\ref{determining-t}), we obtain an alternative
expression of $P_{\rm success}$ with $l^*$ and $L$:
\begin{equation}
 P_{\rm success} = \frac{L}{L-l^*+1} \sum_{i=l^*}^d p_i.
\end{equation}
On the other hand, equation~(\ref{determining-t}) gives $t(l^*-1)<tL$, i.e.,
$l^* \in [1, L]$.
Thus, we have $P_L \leq P_{\rm success}$ due to the right-hand side of
equation~(\ref{Lo}), which means the success probability of our scheme is
optimal because $P_L$ has already been proven optimal.

Therefore, we obtain the following equation that connects the size of a
maximally entangled state $L$, the truncating probability $t$, and the optimal
success probability of concentration $P_L$:
\begin{equation}
 P_L = t L.
\label{revisitedfinal}
\end{equation}

\section{\label{Sec:Types}The method of types}

In order to analyze entanglement concentration in the asymptotic limit,
we will employ the method of types in the following sections.
In this section, we briefly summarize relevant definitions and lemmata on the
method of types without giving the proofs.
For detailed discussions and proofs, see chapter 12 in reference~\cite{CT} and
chapter 1 in reference~\cite{CK}.

Let $X_1, X_2, \cdots, X_n$ be a sequence of $n$ symbols from an alphabet
$A = \{ a_1, a_2, \cdots, a_d \}$, where $d$ is the number of symbols in the
alphabet $A$.
A sequence $x_1x_2 \cdots x_n$ will be denoted by $\mathbf{x}$. 

\begin{Def}
 The type $p_{\mathbf{x}}$ of a sequence $x_1x_2 \cdots x_n$ is the relative
 proportion of occurrences of each symbol of $A$, i.e.,
 \begin{equation}
  p_{\mathbf{x}}(a) = \frac{N(a|\mathbf{x})}{n} \qquad {\rm for \ all}
 \quad a \in A,
 \end{equation}
 where $N(a|\mathbf{x})$ is the number of times the symbol $a$ occurs in the
 sequence $\mathbf{x} \in A^n$.
\end{Def}

A type $p_{\mathbf{x}}$ is a map from symbols $a \in A$ to their frequencies
in the sequence $\mathbf{x}$ (empirical probability distribution).
We denote the set of types with denominator $ n $ by
$\mathcal{P}_{\mathit{n}}$.
If $p \in \mathcal{P}_{\mathit{n}}$, then the set of sequences of length $n$
and type $p$ is called the type class of $p$, denoted by $T_{p}^{n}$, i.e.,
\begin{equation}
 T_{p}^{n} = \{ \mathbf{x} \in A^n| p_{\mathbf{x}} = p \}.
\end{equation}
In other words, $n$-letter sequences that coincide when rearranged in
alphabetical order are in the same type class.

\begin{lem}
 \label{lem:number of types}
 \quad
 \begin{equation}
  | \mathcal{P}_{n} | \leq (n+1)^d. 
 \label{numberoftypes2}
 \end{equation}
\end{lem}

Though the number of sequences in $A^n$ is exponential in $n$, the number of
\textit{types} grows at most polynomially in $n$, which means that at least one
type has exponentially many sequences in its type class.

If each letter in sequences is drawn i.i.d.~ according to some probability
distribution, then we can estimate the probability with which a sequence occurs
by using the Shannon entropy $H(p)$ and the relative entropy
$D(p \parallel q)$, i.e.,
\begin{equation}
 H(p) = -\sum _{i=1}^d p_i \log p_i,
\end{equation}
and
\begin{equation}
 D(p \parallel q) = \sum _{i=1}^d p_i \log \frac{p_i}{q_i},
\end{equation}
where $p$ and $q$ are probability distributions.

\begin{lem}
 \label{lem:prob of seq}
 If $X_1, X_2, \cdots, X_n$ are drawn i.i.d.~according to $q(x)$, then the
 probability of $\mathbf{x}$ depends only on its type and is given by
 \begin{equation}
  q^n ( \mathbf{x}) = 2^{-n \left\{ H(p_{\mathbf{x}}) + D(p_{\mathbf{x}}
  \parallel q) \right\} }.
 \label{probofseq2}
 \end{equation}
\end{lem}

Furthermore, we can also estimate the size of a type class $T_p^n$;
the number of the sequences in $T_p^n$ is bounded as follows:

\begin{lem}
 \label{lem:sizeT}
 For arbitrary type $p \in \mathcal{P}_{n}$,
 \begin{equation}
  \frac{1}{(n+1)^d} 2^{n H(p)} \leq | T_p^n | \leq 2^{n H(p)}.
 \label{sizeT2}
 \end{equation}
\end{lem}

From Lemmata \ref{lem:prob of seq} and \ref{lem:sizeT}, we obtain the bound of
the probability of a type class.

\begin{lem}
 \label{lem:prob of type}
 For arbitrary type $p \in \mathcal{P}_{n}$ and arbitrary probability
 distribution $q$, the probability of the type class $T_p^n$ under $q^n$ is
 bounded as
 \begin{equation}
  \frac{1}{(n+1)^d} 2^{-n D(p \parallel q)} \leq q^n (T_p^n) \leq
  2^{-n D(p \parallel q)}.
 \label{proboftype2}
 \end{equation}
\end{lem}

\begin{figure}[ht]
 \begin{center}
  \includegraphics{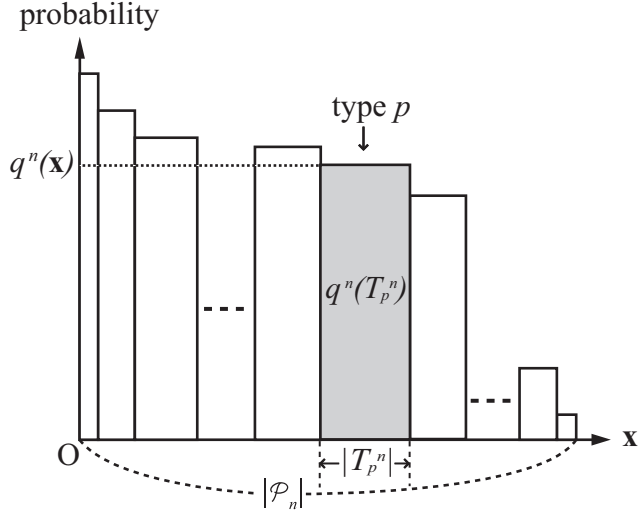}
 \end{center}
 \caption{\label{Fig2} A schematic summary of the method of types.
  This bar graph represents the probability distribution on $n$-letter
  sequences that are drawn i.i.d.~according to $q(x)$.
  Each bar corresponds to a type class, and consists of the sequences of the
  type.
  The number of `stairs' is the number of types $\mathcal{P}_{n}:
  | \mathcal{P}_{n} | \leq (n+1)^d$.
  The height of each bar is the probability with which each sequence in the
  type class occurs:
  $q^n ( \mathbf{x}) = 2^{-n \left\{ H(p_{\mathbf{x}}) + D(p_{\mathbf{x}}
  \parallel q)  \right\} }$.
  The width of each bar is the number of sequences in the type class:
  $(n+1)^{-d} 2^{n H(p)} \leq | T_p^n | \leq 2^{n H(p)}$.
  The area of each bar is the probability of the type class:
  $(n+1)^{-d} 2^{-n D(p \parallel q)} \leq q^n (T_p^n) \leq
  2^{-n D(p \parallel q)}$.
  (Note that if $q(x)$ \textit{degenerates}, i.e., $q(x_i)=q(x_j)$ for some
  $x_i$ and $x_j$ such that $i \neq j$, then sequences in different type
  classes can occur with the same probability.
  Thus, one bar can consist of different type classes.)}
\end{figure}

The lemmata summarized above are also shown schematically in figure~\ref{Fig2}
for easy reference in the proofs of our results, where the method of types
will be heavily used.

\section{\label{Sec:Asymptotic}Asymptotic entanglement concentration}

This section presents the main result of this paper, asymptotic entanglement
concentration from the viewpoint of error exponents.
Suppose we wish to distil a maximally entangled state of size $L_n$ from
$n$ identical copies of $\ket{\phi}$, i.e.,
$\ket{\phi}^{\otimes n} = \sum_{\mathbf{i}} \sqrt{p^{n} (\mathbf{i})}
\ket{\mathbf{i}} \ket{\mathbf{i}}$, where $p^n(\mathbf{i})$ is the
$n$-i.i.d.~extension of $p_i$.
Applying the results in section~\ref{Sec:Revisited} to the $n$-i.i.d. case, we
also have the optimal success probability
\begin{equation}
 P_{L_n} = \sum_{\mathbf{i}} \min\{t_n, p^n( \mathbf{i} ) \},
 \label{P area}
\end{equation}
and the relation between $P_{L_n}$, $L_n$, and the truncating line $t_n$,
\begin{equation}
 P_{L_n} = t_n L_n.
 \label{PvstvsL}
\end{equation}
In the following, we consider the case where the optimal success probability
$P_{L_n}$ converges to one as the number of entangled pairs $n$ increases.
The rate of the convergence is represented by an \textit{error exponent} $r$,
the first-order coefficient in the exponent of the failure probability in the
asymptotic limit, which is defined as
\begin{equation}
 r= \lim_{n \to \infty} \left\{ -\frac{1}{n} \log (1-P_{L_n}) \right\}.
\end{equation}
Intuitively, this means that the error probability behaves as $2^{-nr}$.

We will derive the maximum number of Bell pairs distilled per copy in the
asymptotic limit, $E$, as a function of the error exponent $r$.
First, we prove a theorem that relates entanglement yield and an error exponent
via a monotone function, from which we will derive a formula for entanglement
yield $E(r)$.

\begin{thm}
 \label{thm:Direct}
 Consider a sequence of entanglement concentration schemes converting $n$
 identical copies of $\ket{\phi} = \sum_{i=1}^{d} \sqrt{p_i} \ket{i} \ket{i}$,
 i.e., $\ket{\phi}^{\otimes n}$, into a maximally entangled state of size
 $L_n$, which attain the optimal success probability $P_{L_n}$.
 Suppose
 \begin{equation}
  \limsup_{n \to \infty} \left( \frac{1}{n} \log L_n \right) < H(p),
  \label{less than H}
 \end{equation}
 and
 \begin{equation}
  \frac{1}{n} \log L_n > -\log p_1,
  \label{more than -log p1}
 \end{equation}
 where $p=(p_1, \cdots, p_d)$.
 Then, 
 \begin{equation}
   \limsup_{n \to \infty} \left( \frac{1}{n}\log L_n \right)
  = f \left( \liminf_{n \to \infty} \left\{ -\frac{1}{n} \log (1-P_{L_n})
    \right\} \right ),
  \label{liminf1}
 \end{equation}
 and
 \begin{equation}
   \liminf_{n \to \infty} \left( \frac{1}{n}\log L_n \right)
  = f \left( \limsup_{n \to \infty} \left\{ -\frac{1}{n} \log (1-P_{L_n})
   \right\} \right),
  \label{limsup1}
 \end{equation}
 where
 \begin{equation}
  f(r) \equiv \min_{q:D(q \parallel p) \leq r} \left\{ D(q \parallel p)+H(q)
  \right\}.
 \label{f} 
 \end{equation}
\end{thm}

\begin{figure}[ht]
 \begin{center}
  \includegraphics{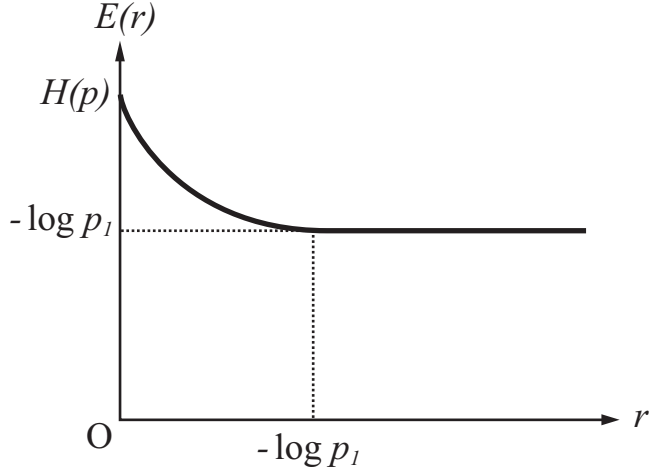}
 \end{center}
 \caption{\label{Fig3} Entanglement yield in asymptotic entanglement
  concentration with an error exponent $r$.
  The horizontal axis represents the error exponent.
  The vertical axis represents the number of Bell pairs distilled per copy in
  the asymptotic limit:
  $E(r) = \min_{q:D(q \parallel p) \leq r} \left\{ D(q \parallel p) + H(q)
  \right\}$.}
\end{figure}

\begin{prf}
 Let
 \begin{equation}
  R_n \equiv -\frac{1}{n} \log t_n.
  \label{R and t}
 \end{equation}
 Then, equation~(\ref{PvstvsL})gives
 \begin{equation}
  R_n = \frac{1}{n} \log L_n - \frac{1}{n} \log P_{L_n}.
  \label{Rn Ln and PLn}
 \end{equation}
 
 In what follows, we only consider a convergent sub-sequence of $\{R_n\}$,
 that is, take an infinite subset $\mathcal{N} \subset \{1,2,\ldots \}$ such
 that $R' \equiv \lim_{n \to \infty, n \in \mathcal{N}} R_n$ exists.
 For simplicity, we denote the sub-sequence $\{ R_n \}_{n \in \mathcal{N}}$
 as  $\{ R'_n \}$, and omit $n \in \mathcal{N}$.
 Since equation~(\ref{less than H}) implies $\lim_{n \to \infty} P_{L_n} = 1$,
 equation~(\ref{Rn Ln and PLn}) gives
 \begin{equation}
  \lim_{n \to \infty} \left( \frac{1}{n} \log L'_n \right) = R',
  \label{lim L}
 \end{equation}
 where $\{ L'_n \}$ denotes the sub-sequence that corresponds to $\{ R'_n \}$.
 Then, equations~(\ref{less than H}) and (\ref{more than -log p1}) imply
 \begin{equation}
  -\log p_1 \leq R' < H(p).
  \label{conditions on R'}
 \end{equation}
 Equation~(\ref{P area}) gives
 \begin{equation}
  -\frac{1}{n} \log ( 1- P_{L'_n}) =
  -\frac{1}{n} \log \left( 1- \sum_{\mathbf{i}} \min\{t'_n, p^n(\mathbf{i}) \}
   \right),
  \label{error prob log}
 \end{equation}
 where $\{t'_n \}$ is a sub-sequence such that $R'_n = - n^{-1} \log t'_n$.
 
 In the following, we estimate the right-hand side of
 equation~(\ref{error prob log}) by the method of types.
 Rewriting the area above the truncating probability $t'_n$ in terms of the
 type theory, we obtain
 \begin{equation}
  1- \sum_{\mathbf{i}} \min \{ t'_n, p^n(\mathbf{i}) \} =
  \sum_{q \in \mathcal{P}_{n} : p^n(q) \geq t'_n}
  \left( p^n(T_q ^n) - | T_q ^n | t'_n \right).
  \label{error prob area}
 \end{equation}
 Invoking equations~(\ref{numberoftypes2}), (\ref{probofseq2}),
 and(\ref{proboftype2}), we have the inequalities
 \begin{eqnarray}
 \fl \sum_{q \in \mathcal{P}_{n} : p^n(q) \geq t'_n} \left( p^n(T_q ^n) -
  | T_q ^n | t'_n \right) 
  &\leq& \sum_{q \in \mathcal{P}_{n} : p^n(q) \geq t'_n} p^n(T_q ^n)
         \nonumber \\
  &\leq& (n+1)^d \max_{q \in \mathcal{P}_{n} :D(q \parallel p)+ H(q) \leq R'_n}
         2^{-n D(q \parallel p)}.
 \end{eqnarray}
 Note that $p^n(q) \geq t'_n$ is equivalent to
 $D(q \parallel p) + H(q) \leq R'_n$ due to equations~(\ref{probofseq2}) and
 (\ref{R and t}).
 Thus, together with equations~(\ref{error prob log}) and
 (\ref{error prob area}), we have
 \begin{equation}
 \fl \qquad -\frac{1}{n} \log ( 1- P_{L'_n}) 
    \geq  -\frac{d}{n} \log (n+1) + \min_{q \in \mathcal{P}_{n} :
         D(q \parallel p)+ H(q) \leq R'_n} D(q \parallel p).
  \label{pre liminf1}
 \end{equation}
 Therefore, for $-\log p_1 \leq R' < H(p)$,
 \begin{equation}
  \liminf_{n \to \infty} \left\{ -\frac{1}{n} \log ( 1- P_{L'_n}) \right\} \geq
  \min_{q :D(q \parallel p)+ H(q) \leq R'} D(q \parallel p).
 \label{liminf1'}
 \end{equation}
 
 Next, first we consider the case $-\log p_1 < R'$.
 Then, for any $q$ satisfying
 \begin{equation}
  -\log p_1 \leq D(q \parallel p)+ H(q) < R',
  \label{q and R'}
 \end{equation}
 there exists a sequence of types $q'_n \in \mathcal{P}_{n}$ such that
 \begin{equation}
  D(q'_n \parallel p)+ H(q'_n) \leq R'_n \qquad {\rm with} \qquad
   \lim_{n \to \infty} q'_n =q.
  \label{lim q'_n=q}
 \end{equation}
 Invoking equations~(\ref{sizeT2}) and (\ref{proboftype2}), we also have
 \begin{eqnarray}
 \fl \sum_{q \in \mathcal{P}_{n} : p^n(q) \geq t'_n} \left(p^n(T_q ^n) -
     | T_q ^n | t'_n \right)
  &\geq& p^n(T_{q_n'} ^n) - | T_{q_n'} ^n | t'_n \nonumber \\
  &\geq& \frac{1}{(n+1)^d} 2^{-nD(q_n' \parallel p)} - 2^{n H(q_n')}2^{-nR'_n}.
 \end{eqnarray}
 Thus, together with equations~(\ref{error prob log}) and
 (\ref{error prob area}), we have
 \begin{equation}
 \fl \qquad -\frac{1}{n} \log ( 1- P_{L'_n}) \leq -\frac{1}{n} \log \left\{
     2^{-n \left( D(q_n' \parallel p) + \frac{d}{n} \log(n+1)
     \right)} -  2^{-n (R'_n - H(q_n'))} \right\}.
  \label{before limsup1}
 \end{equation}
 Equations~(\ref{q and R'}) and (\ref{lim q'_n=q}) imply
 \begin{equation}
  \lim_{n \to \infty} \left\{ D(q_n' \parallel p) + \frac{d}{n} \log(n+1)
  \right\} < \lim_{n \to \infty} \left( R'_n - H(q_n') \right).
 \end{equation}
 Applying Lemma \ref{lem:limit} in \ref{App:greater exponent} to
 equation~(\ref{before limsup1}), we obtain
 \begin{equation}
  \limsup_{n \to \infty} \left\{ -\frac{1}{n} \log (1 -P_{L'_n}) \right\}
  \leq D(q \parallel p),
  \label{pre limsup1}
 \end{equation}
 which holds for any $q$ satisfying equation.~(\ref{q and R'}).
 Therefore, for $-\log p_1 < R' <H(p)$,
 \begin{eqnarray}
  \limsup_{n \to \infty} \left\{ -\frac{1}{n} \log (1-P_{L'_n}) \right\}
  &\leq& \inf_{q :D(q \parallel p)+ H(q) < R'} D(q \parallel p) \nonumber \\
  &=& \min_{q :D(q \parallel p)+ H(q) \leq R'} D(q \parallel p).
  \label{limsup1'}
 \end{eqnarray}
 
 Equations~(\ref{liminf1'}) and (\ref{limsup1'}) show that when
 $-\log p_1 < R' <H(p)$, the sub-sequence $\left\{ - n^{-1}
 \log (1-P_{L'_n}) \right\}$ is also convergent and
 \begin{equation}
  r' \equiv \lim_{n \to \infty} \left\{ -\frac{1}{n} \log (1-P_{L'_n}) \right\}
  = \min_{q :D(q \parallel p)+ H(q) \leq R'} D(q \parallel p).
  \label{r' and R'}
 \end{equation}
 According to Corollary~\ref{cor:newmonotonicity2} in
 \ref{App:NewMonotonicity},
 setting $S(q) \equiv D(q \parallel p) + H(q)$, $U(q) \equiv D(q \parallel p)$,
 $x \equiv R$, and noting that $x_1=-\log p_1$, and $x_2 = H(p)$, we see that
 the function $ R \mapsto
 \min_{q:D(q \parallel p) + H(q) =R} D(q \parallel p)$ is
 continuous and strictly monotone decreasing in $(0, -\log p_1)$ for
 $R \in ( -\log p_1, H(p))$.
 Thus, so is the function $ R \mapsto
 \min_{q:D(q \parallel p) + H(q) \leq R} D(q \parallel p)$,
 which is the inverse function of $f(r)$ [equation~(\ref{f})]
 (see figure~\ref{Fig3}).
 Therefore, equations~(\ref{lim L}) and (\ref{r' and R'}) provide
 \begin{equation}
  \lim_{n \to \infty} \left\{ -\frac{1}{n} \log (1-P_{L'_n}) \right\}
  = f^{-1} \left( \lim_{n \to \infty} \left( \frac{1}{n} \log L'_n \right)
  \right),
 \end{equation}
 i.e.,
 \begin{equation}
  \lim_{n \to \infty} \left( \frac{1}{n} \log L'_n \right)
  = f \left( \lim_{n \to \infty} \left\{ -\frac{1}{n} \log (1-P_{L'_n})
  \right\} \right),
  \label{R' and r'}
 \end{equation}
 for
 \begin{equation}
  0 <\lim_{n \to \infty} \left\{ -\frac{1}{n} \log (1-P_{L'_n}) \right\}
  <-\log p_1,
  \label{rage of r'}
 \end{equation}
 and
 \begin{equation}
  -\log p_1 <\lim_{n \to \infty} \left( \frac{1}{n} \log L'_n \right)<H(p).
  \label{rage of R'}
 \end{equation}
  
 On the other hand, when $R'=- \log p_1$, equations~(\ref{lim L}) and
 (\ref{liminf1'}) give
 \begin{equation}
  \lim_{n \to \infty} \left(\frac{1}{n} \log L'_n \right)= -\log p_1,
 \end{equation}
 and
 \begin{equation}
  \liminf_{n \to \infty} \left\{ -\frac{1}{n} \log (1-P_{L'_n}) \right\}
  \geq -\log p_1.
 \end{equation}
 In addition, for $r \geq -\log p_1$,
 \begin{equation}
  f(r)=\min_{q:D(q \parallel p) \leq r} \left\{ D(q \parallel p)+H(q) \right\}
  = - \log p_1.
 \end{equation}
 Note that $D(q \parallel p) + H(q) = -\sum_{i=1}^{d} q_i \log p_i \geq
 -\log p_1$ with equality when $q=(1,0,\cdots,0)$, i.e., $D(q \parallel p) =
 -\log p_1$.
 Hence,
 \begin{eqnarray}
 \fl \qquad -\log p_1 = \lim_{n \to \infty} \left(\frac{1}{n} \log L'_n \right)
    &=& f \left( \liminf_{n \to \infty} \left\{ -\frac{1}{n} \log (1-P_{L'_n})
        \right\} \right) \nonumber \\
    &=& f \left( \limsup_{n \to \infty} \left\{ -\frac{1}{n} \log (1-P_{L'_n})
        \right\} \right),
  \label{E=-log p1}
 \end{eqnarray} 
 for
 \begin{equation}
  \liminf_{n \to \infty} \left\{ -\frac{1}{n} \log (1-P_{L'_n}) \right\}
  \geq -\log p_1.
  \label{liminf r}
 \end{equation}
  
 The above argument holds for any convergent sub-sequences $\{ R'_n \}$, and
 $f(r)$ is monotone decreasing.
 Therefore, equations~(\ref{R' and r'}), (\ref{rage of r'}),
 (\ref{rage of R'}), (\ref{E=-log p1}), and (\ref{liminf r}) provide equations~
 (\ref{liminf1}) and (\ref{limsup1})
 (see figure~\ref{Fig3}).
\end{prf}

Theorem \ref{thm:Direct} leads to the following corollary, which gives the
maximum asymptotic entanglement yield $E(r)$ under the requirement that the
failure probability decreases as rapidly as $2^{-nr}$:

\begin{cor}
 \label{cor:Direct cor}
 Consider a sequence of entanglement concentration schemes converting
 $\ket{\phi}^{\otimes n}$ into a maximally entangled state of size $L_n$ with
 success probability $P_{\rm success}^{(n)}$, such that
\begin{equation}
 r \leq \liminf_{n \to \infty} \left\{ -\frac{1}{n} \log
 (1-P_{\rm success}^{(n)}) \right\}.
\end{equation}
Let us denote the class of all such sequences by $\mathcal{C} (r)$.
Then, for $r>0$,
\begin{eqnarray}
 E(r) &\equiv& \max_{\mathcal{C} (r)} \limsup_{n \to \infty} \left( \frac{1}{n}
 \log L_n \right) = \max_{\mathcal{C} (r)} \liminf_{n \to \infty} \left(
 \frac{1}{n} \log L_n \right) \nonumber \\
 &=& \min_{q:D(q \parallel p) \leq r} \left\{D(q \parallel p) + H(q) \right\}.
 \label{direct}
\end{eqnarray}
\end{cor}

This corollary connects the following two facts on the distillable entanglement
of bipartite pure states $E$:
\begin{enumerate}
 \item If we allow error probability that vanishes in the asymptotic limit, $E$
  cannot exceed $H(p)$ \cite{Bennett96}.
 \item If we stick to deterministic strategy even in the finite regimes (i.e.,
  no error is allowed), $E$ is equal to $- \log p_1$ \cite{Morikoshi2001}.
\end{enumerate}
Equation~(\ref{direct}) provides the missing link between them, i.e.,
$E_{\rm entropy} = H(p) = \lim_{r \to 0} E(r)$ and
$E_{\rm det}  = -\log p_1 = \lim_{r \to \infty} E(r)$.
The distillable entanglement $E(r)$ monotonically decreases and reaches
$- \log p_1$ when $r = -\log p_1$, which means that probabilistic concentration
schemes with error exponents greater than $-\log p_1$ effectively give the
same result as deterministic ones.

Next, we move to a discussion about entanglement yield and fidelity.
Suppose we wish to transform the initial state $\ket{\phi}^{\otimes n}$
deterministically into some final (possibly mixed) state that is as close to a
maximally entangled state $\ket{\Phi_{L_n}}$ of size $L_n$ as possible.
Instead of the success probability $P_{success}^{(n)}$ in the previous
argument, we here require that the fidelity $F_n$ between the final state and
the maximally entangled state $\ket{\Phi_{L_n}}$ approach unity as rapidly as
$1-2^{-nr}$, namely,
\begin{equation} 
 r \leq \liminf_{n \to \infty} \left\{ -\frac{1}{n} \log (1-F_n) \right\}.
\end{equation}
Let us denote the class of all such sequences by $\mathcal{C}_F(r)$.
The maximum asymptotic entanglement yield $E_F(r)$ over $\mathcal{C}_F(r)$ can
be obtained by reducing the problem to that of probabilistic schemes via the
following two lemmata.
\begin{lem}
 \label{lem:lemma:0}
 If the transformation $\ket{\phi} \longrightarrow \ket{\Phi_L}$ is possible
 with probability $1-\epsilon$, then there exists a deterministic
 transformation $\ket{\phi} \longrightarrow \rho$ with fidelity
 $\bra{\Phi_L}\rho\ket{\Phi_L} \geq 1-\epsilon$.
\end{lem}
The proof is straightforward by considering the case
$\rho=(1-\epsilon)\ket{\Phi_L}\bra{\Phi_L}+\epsilon \rho'$.
This lemma implies that for any sequence of probabilistic schemes belonging to
$\mathcal{C} (r)$ with size $\{L_n\}$, there exists a sequence belonging to
$\mathcal{C}_F(r)$ with the same size $\{L_n\}$.
Hence,
\begin{equation}
 \max_{\mathcal{C}_F(r)} \liminf_{n \to \infty} \left(\frac{1}{n} \log L_n
 \right) \geq E(r).
 \label{prob < fidelity:revised}
\end{equation}

\begin{lem}
 \label{lem:lemma:1}
 If there exists a deterministic transformation $\ket{\phi} \longrightarrow
 \rho$ with fidelity $\bra{\Phi_T}\rho\ket{\Phi_T} \geq 1-\epsilon$, the
 transformation $\ket{\phi} \longrightarrow \ket{\Phi_L}$ is possible with
 probability $1-6\epsilon$, for $L=\left\lfloor T(1-6\epsilon)/6
 \right\rfloor$.
\end{lem}
\begin{prf}
 See \ref{App:Fidelity lemmata}.
\end{prf}
This lemma implies that for any sequence of deterministic schemes belonging to
$\mathcal{C}_F(r)$ with size $\{L_n\}$, there exists a sequence belonging to
$\mathcal{C} (r)$ with the size $\{L'_n\}$, where $L'_n \geq L_n / 7$
for large $n$.
Hence,
\begin{equation}
  \max_{\mathcal{C}_F(r)} \limsup_{n \to \infty} \left(\frac{1}{n} \log L_n
 \right) \leq \max_{\mathcal{C} (r)} \limsup_{n \to \infty} \left\{ \frac{1}{n}
 \log (7 L'_n) \right\} = E(r).
 \label{fidelity < prob:revised}
\end{equation}
From Eqs~(\ref{prob < fidelity:revised}) and (\ref{fidelity < prob:revised}),
we have 
\begin{equation}
\fl \qquad E_F(r) \equiv \max_{\mathcal{C}_F(r)} \limsup_{n \to \infty}
     \left(\frac{1}{n} \log L_n \right)=\max_{\mathcal{C}_F(r)}(r),
 \label{directfidelityyield}
     \liminf_{n \to \infty} \left(\frac{1}{n} \log L_n\right)=E
\end{equation}
which shows that the entanglement yield $E_F$ can be expressed as the same
function as that for probabilistic cases.
Though the probabilistic argument assumes that we can transform the initial
state to the exact copy of a maximally entangled state with probability
close to one, the fidelity argument is much more natural in that it imposes
weaker restrictions on manipulation where the fidelity close to one,
which is common to more general entanglement manipulations,
such as entanglement dilution.

\section{\label{Sec:Strong}Strong converse}

We have investigated how entanglement yield behaves when the failure
probability exponentially decreases.
In this section, conversely, we discuss asymptotic entanglement concentration
with exponentially decreasing \textit{success} probability, which
will finally lead to the strong converse of asymptotic entanglement
concentration.

The following discussion is parallel to that in the previous section.
So, when we distil a maximally entangled state of size $L_n^*$ from
$\ket{\phi}^{\otimes n}$, the following relations also hold:
\begin{equation}
 P_{L_n^*} = \sum_{\mathbf{i}} \min\{t_n, p^n( \mathbf{i} ) \},
 \label{Pvsr2}
\end{equation}
and
\begin{equation}
 P_{L_n^*} = t_n L_n^*.
 \label{PvstvsL2}
\end{equation}

We assume that the success probability converges to zero as the number of
the entangled pairs $n$ increases.
Then, the first-order coefficient in the exponent of the success probability
in the asymptotic limit becomes
\begin{equation}
 r = \lim_{n \to \infty} \left( -\frac{1}{n} \log P_{L_n^*} \right).
\end{equation}
Intuitively, this means that the success probability behaves as $2^{-nr}$.
We will derive the maximum number of Bell pairs distilled per copy in the
asymptotic limit, $E^*$, as a function of the exponent $r$.
First, we prove a theorem that relates entanglement yield and an exponent via a
monotone function, from which we will derive a formula for entanglement yield
$E^*(r)$.

\begin{thm}
 \label{thm:Converse}
 Consider a sequence of entanglement concentration schemes that convert $n$
 identical copies of $\ket{\phi} = \sum_{i=1}^{d} \sqrt{p_i} \ket{i} \ket{i}$,
 i.e., $\ket{\phi}^{\otimes n}$, into a maximally entangled state of size
 $L_n^*$, which attain the optimal success probability $P_{L_n^*}$.
 Suppose
 \begin{equation}
  \liminf_{n \to \infty} \left( \frac{1}{n} \log L_n^* \right) > H(p),
  \label{more than H}
 \end{equation}
 and
 \begin{equation}
  \frac{1}{n} \log L_n^* < \log d,
  \label{less than log d}
 \end{equation}
 where $p=(p_1, \cdots p_d)$.
 Then,
 \begin{equation}
   \limsup_{n \to \infty} \left( \frac{1}{n} \log L_n^* \right) =
  g\left(\limsup_{n \to \infty} \left( -\frac{1}{n} \log P_{L_n^*} \right)
  \right),
  \label{limsup2}
 \end{equation}
 and
 \begin{equation}
   \liminf_{n \to \infty} \left( \frac{1}{n} \log L_n^* \right) =
  g\left(\liminf_{n \to \infty}  \left( -\frac{1}{n} \log P_{L_n^*} \right)
  \right),
  \label{liminf2}
 \end{equation}
 where
 \begin{equation}
  g(r) \equiv \max_{q:D(q \parallel p) \leq r} H(q).
 \end{equation}
\end{thm}

\begin{prf}
 Let
 \begin{equation}
  R_n \equiv -\frac{1}{n} \log t_n.
  \label{Rt*}
 \end{equation}
 Then, equation~(\ref{PvstvsL2}) gives
 \begin{equation}
  -\frac{1}{n} \log P_{L_n^*} = R_n -\frac{1}{n} \log L_n^*.
  \label{limPLn*}
 \end{equation}
 In what follows, we only consider a convergent sub-sequence of $\{R_n\}$,
 that is, take an infinite subset $\mathcal{N} \subset \{1,2,\ldots \}$ such
 that $R' \equiv \lim_{n \to \infty, n \in \mathcal{N}} R_n$ exists.
 For simplicity, we denote the sub-sequence $\{ R_n \}_{n \in \mathcal{N}}$
 as  $\{ R'_n \}$, and omit $n \in \mathcal{N}$.
 Equations~(\ref{more than H}) and (\ref{limPLn*}) imply
 \begin{equation}
  H(p) < R'.
  \label{<R'}
 \end{equation}
 First, we consider the case
 \begin{equation}
  R' < -\frac{1}{d}\sum_{i=1}^d \log p_i.
  \label{R'<}
 \end{equation}
 Then, for sufficiently large $n$, we have
 \begin{equation}
  R'_n < -\frac{1}{d}\sum_{i=1}^d \log p_i.
  \label{Rn'<}
 \end{equation}
 Equation~(\ref{Pvsr2}) gives
 \begin{equation}
  -\frac{1}{n} \log P_{L_n^{*'}} = -\frac{1}{n} \log \sum_{\mathbf{i}}
  \min\{t'_n, p^n(\mathbf{i}) \},
  \label{area2}
 \end{equation}
 where $\{P_{L_n^{*'}} \}$ and $\{t'_n\}$ are sub-sequences related to
 $\{R'_n \}$ by equations~(\ref{Rt*}) and (\ref{limPLn*}).
 
 In the following, we estimate the right-hand side of equation~(\ref{area2}) by
 the method of types.
 Rewriting the area below the truncating probability $t'_n$ in terms of the
 type theory, we obtain
 \begin{equation}
 \fl \qquad \sum_{\mathbf{i}} \min\{ t'_n, p^n(\mathbf{i}) \} =
     \sum_{q \in \mathcal{P}_{n} : p^n(q) > t'_n}| T_q ^n | t'_n +
     \sum_{q \in \mathcal{P}_{n} : p^n(q) \leq t'_n} p^n(T_q ^n).
  \label{area3}
 \end{equation}
 Invoking equations~(\ref{numberoftypes2}), (\ref{probofseq2}),
 (\ref{sizeT2}), and (\ref{proboftype2}),
 we have the inequalities
 \begin{eqnarray}
 \fl  \sum_{q \in \mathcal{P}_{n} : p^n(q) > t'_n} | T_q ^n | t'_n
      +\sum_{q \in \mathcal{P}_{n} : p^n(q) \leq t'_n} p^n(T_q ^n)
      \nonumber \\
 \fl \leq \sum_{q \in \mathcal{P}_{n} : D(q \parallel p) + H(q) < R'_n}
         2^{n(H(q)-R'_n)}+\sum_{q \in \mathcal{P}_{n} :
         D(q \parallel p) + H(q) \geq R'_n} 2^{-nD(q \parallel p)} \\
 \fl \leq (n+1)^d \left[ \max_{q \in \mathcal{P}_{n} : D(q \parallel p) + H(q)
         \leq R'_n} 2^{n(H(q)-R'_n)} +
         \max_{q \in \mathcal{P}_{n} : D(q \parallel p) + H(q)
         \geq R'_n} 2^{-nD(q \parallel p)} \right].
  \label{pre liminf2}
 \end{eqnarray}
 
 According to Lemma~\ref{lem:newmonotonicity} in
 \ref{App:NewMonotonicity},
 setting $S(q) \equiv D(q \parallel p) + H(q)$, $T(q) \equiv H(q)$,
 $x \equiv R$, and noting that $x_1= -\log p_1 $,
 and $x_2 = -d^{-1} \sum_{i=1}^d \log p_i$,
 we see that the function $ R \mapsto
 \max_{q:D(q \parallel p) + H(q) =R} H(q)$ is continuous and strictly monotone
 increasing in $[0, \log d]$ for $R \in \left[-\log p_1, -d^{-1}
 \sum_{i=1}^d \log p_i \right]$. 
 Thus, $\max_{q \in \mathcal{P}_n :D(q \parallel p) + H(q) \leq R'_n} H(q)
 = \max_{q \in \mathcal{P}_n :D(q \parallel p) + H(q) = R'_n} H(q) \\
 = \max_{q \in \mathcal{P}_n :D(q \parallel p) + H(q) = R'_n}
   [R'_n -D(q \parallel p)]$.
 
 Similarly, according to Corollary~\ref{cor:newmonotonicity2} in
 \ref{App:NewMonotonicity},
 setting $S(q) \equiv - \{ D(q \parallel p) + H(q)\}$,
 $U(q) \equiv D(q \parallel p)$, $x \equiv -R$, 
 and noting that $x_1= \log p_d$, and $x_2 = -H(p)$,
 we see that the function $ -R \mapsto
 \min_{q:D(q \parallel p) + H(q) = R} D(q \parallel p)$ is
 continuous and strictly monotone decreasing in $[0, -\log p_d]$ for
 $-R \in [\log p_d, -H(p)]$.
 This means that the function $ R \mapsto
 \min_{q:D(q \parallel p) + H(q) =R} D(q \parallel p)$ is
 continuous and strictly monotone increasing in $[0, -\log p_d]$ for
 $R \in [H(p),-\log p_d]$.
 Thus, $\min_{q \in \mathcal{P}_n :D(q \parallel p) + H(q) \geq R'_n}
 D(q \parallel p) = \min_{q \in \mathcal{P}_n :D(q \parallel p) + H(q) = R'_n}
 D(q \parallel p)$.
 
 Then, combining equations~(\ref{area3}) and (\ref{pre liminf2}) and using
 equation~(\ref{numberoftypes2}), we have
 \begin{equation}
  \sum_{\mathbf{i}} \min\{ t'_n, p^n(\mathbf{i}) \} \leq 2(n+1)^d
   \max_{q \in \mathcal{P}_n:
      D(q \parallel p) + H(q) = R'_n} 2^{-nD(q \parallel p)}.
 \end{equation}
 Together with equation~(\ref{area2}),
 we obtain
 \begin{equation}
  \liminf_{n \to \infty} \left( -\frac{1}{n} \log P_{L_n^{*'}} \right)
  \geq \min_{q: D(q \parallel p) + H(q) = R'}  D(q \parallel p).
  \label{liminf3}
 \end{equation}
 
 Next, we derive a lower bound of equation~(\ref{area3}).
 For any $q$ satisfying $D(q \parallel p) + H(q) = R' $, there exists a
 sequence of  types $q_n' \in \mathcal{P}_{n}$ such that $D(q_n' \parallel p)
 + H(q_n') \geq R'_n$ and $\lim_{n \to \infty} q_n' = q$, due to the condition
 $R'_n < - d^{-1} \sum_{i=1}^{d} \log p_i \leq -\log p_d$.
 
 Invoking equation~(\ref{proboftype2}), we have
 \begin{eqnarray}
 \fl \sum_{q \in \mathcal{P}_{n} : p^n(q) > t'_n} | T_q ^n | t'_n
  + \sum_{q \in \mathcal{P}_{n} : p^n(q) \leq t'_n} p^n(T_q ^n) 
  &\geq& \sum_{q \in \mathcal{P}_{n} : p^n(q) \leq t'_n} p^n(T_q ^n)
         \nonumber \\
  &\geq& p^n(T_{q'_n}^n) \nonumber \\
  &\geq& \frac{1}{(n+1)^d} 2^{-nD(q_n' \parallel p)}.
 \end{eqnarray}
 Together with equations~(\ref{area2}) and (\ref{area3}), we obtain
 \begin{equation}
  \limsup_{n \to \infty} \left( -\frac{1}{n} \log P_{L_n^{*'}} \right)
  \leq D(q \parallel p),
 \end{equation}
 which holds for any $q$ satisfying $D(q \parallel p) + H(q) = R'$.
 Therefore,
 \begin{equation}
  \limsup_{n \to \infty} \left( -\frac{1}{n} \log P_{L_n^{*'}} \right)
  \leq \min_{q: D(q \parallel p) + H(q) = R'} D(q \parallel p).
  \label{limsup3}
 \end{equation}
 
 Equations~(\ref{liminf3}) and (\ref{limsup3}) imply that the sub-sequence
 $\left\{ -n^{-1} \log P_{L_n^{*'}} \right\}$ is also convergent, i.e.,
 \begin{equation}
  r' \equiv \lim_{n \to \infty} \left( -\frac{1}{n} \log P_{L_n^{*'}} \right)
  = \min_{q: D(q \parallel p) + H(q) = R'}  D(q \parallel p).
  \label{r' vs R'}
 \end{equation}
 As stated above, the function $ R \mapsto
 \min_{q:D(q \parallel p) + H(q) =R} D(q \parallel p)$ is
 continuous and strictly monotone increasing in $[0, -\log p_d]$ for
 $R \in [H(p), -\log p_d]$.
 Hence, the inverse function of equation~(\ref{r' vs R'}) clearly exists
 for $R\in(H(p), -\log p_d)$.
 However, now we consider the range restricted to equation~(\ref{R'<}),
 we obtain, for $r\in(0,c)$ with
 $c \equiv D(u \parallel p)$ being the relative entropy between the uniform
 distribution $u=(1/d, \cdots, 1/d)$ and $p$,
 \begin{equation}
  R'(r') = \max_{q:D(q \parallel p) = r'} \left\{ D(q \parallel p) + H(q)
  \right\}.
 \end{equation}
 
 Furthermore, from equation~(\ref{limPLn*}), we see that if $R'$ and $r'$
 exist, then the sub-sequence $\left\{ n^{-1} \log L_n^{*'} \right\}$ is
 also convergent, i.e., $E^{*'} \equiv \lim_{n \to \infty} \left(
 n^{-1} \log L_n^{*'} \right)$ exists, and
 \begin{equation}
  r' = R' - E^{*'}.
 \end{equation}
 Hence,
 \begin{eqnarray}
  E^{*'} = R'(r') -r' &=& \max_{q:D(q \parallel p) = r'}
   \left\{ D(q \parallel p) + H(q) \right\}-r' \nonumber \\
  &=& \max_{q:D(q \parallel p) = r'} H(q).
 \end{eqnarray}
 According to Lemma~\ref{lem:newmonotonicity} in
 \ref{App:NewMonotonicity},
 setting $S(q) \equiv D(q \parallel p)$, $T(q) \equiv H(q)$,
 $x \equiv r$, and noting that $x_1= 0$, and $x_2 = c$,
 we see that the function $ r \mapsto
 \max_{q:D(q \parallel p) =r} H(q)$ is continuous and strictly monotone
 increasing in $(H(p), \log d)$ for $r \in (0,c)$, thus
 \begin{equation}
  \max_{q:D(q \parallel p) = r} H(q) =
  \max_{q:D(q \parallel p) \leq r} H(q).
 \end{equation}
 Therefore, in the case of equation~(\ref{R'<}), we have $0<r'<c$,
 $H(p)<E^{*'}<\log d$, and
 \begin{equation}
  E^{*'}(r') = g(r').
  \label{subsequence2}
 \end{equation}
  
 Next, we consider the case $R' \geq -d^{-1} \sum_{i=1}^{d} \log p_i$.
 equation~(\ref{less than log d}) provides
 \begin{equation}
  \limsup_{n \to \infty} \left( \frac{1}{n} \log L_n^{*'} \right)
  \leq \log d.
  \label{limsup<logd}
 \end{equation}
 Together with equation~(\ref{limPLn*}), we have
 \begin{equation}
 \fl \qquad
     \liminf_{n \to \infty} \left( -\frac{1}{n} \log P_{L_n^{*'}} \right) =
     R'- \limsup_{n \to \infty} \left( \frac{1}{n} \log L_n^{*'} \right) \geq
     R' - \log d.
 \end{equation}
 Hence,
 \begin{equation}
  \liminf_{n \to \infty} \left( -\frac{1}{n} \log P_{L_n^{*'}} \right) \geq
  -\frac{1}{d} \sum_{i=1}^{d} \log p_i -\log d =c.
 \end{equation}
 Then, since the function $E^{*'}(r')$ is monotone increasing, we have \\
 $\liminf_{n \to \infty} \left( n^{-1} \log L_n^{*'} \right) \geq
 \lim_{r' \uparrow c}E^{*'}(r') = g(c)=\log d$.
 Note that for $r \geq c \\ = D(u \parallel p)$,
 $g(r) = \max_{q:D(q \parallel p) \leq r} H(q) = \log d$.
 Hence, together with equation~(\ref{limsup<logd}) we have
 \begin{eqnarray}
  \log d = \lim_{n \to \infty} \left( \frac{1}{n} \log L_n^{*'} \right)
  &=& g \left( \liminf_{n \to \infty} \left( -\frac{1}{n} \log P_{L_n^{*'}}
     \right) \right) \nonumber \\
  &=& g\left( \limsup_{n \to \infty} \left( -\frac{1}{n} \log P_{L_n^{*'}}
     \right) \right),
  \label{liminf = limsup = log d}
 \end{eqnarray}
 for
 \begin{equation}
  \liminf_{n \to \infty} \left( -\frac{1}{n} \log P_{L_n^{*'}} \right) \geq c
  \label{liminf > c}.
 \end{equation}
 
 The above argument holds for any convergent sub-sequence $\{ R'_n \}$.
 Since $r$ is an exponent of success probability, the function $E^{*}(r)$ is
 monotone increasing.
 And, so is $g(r)$.
 Therefore, equations~(\ref{subsequence2}), (\ref{liminf = limsup = log d}),
 and (\ref{liminf > c}) provide equations~(\ref{limsup2}) and (\ref{liminf2}).
\end{prf}

Theorem \ref{thm:Converse} leads to the following corollary that gives the
maximum asymptotic entanglement yield $E^* (r)$ under the requirement that the
success probability decreases as slowly as $2^{-nr}$:

\begin{cor}
 \label{cor:Converse2}
 Consider a sequence of entanglement concentration schemes converting
 $\ket{\phi}^{\otimes n}$ into a maximally entangled state of size $L_n^*$ with
 success probability $P_{\rm success}^{(n)}$, such that
\begin{equation}
 r \geq \limsup_{n \to \infty} \left( -\frac{1}{n} \log
 P_{\rm success}^{(n)} \right).
\end{equation}
Let us denote the class of all such sequences by $\mathcal{C}^* (r)$.
Then, for $r>0$,
\begin{eqnarray}
 E^*(r) &\equiv& \max_{\mathcal{C}^* (r)} \limsup_{n \to \infty} \left(
 \frac{1}{n} \log L_n^* \right) = \max_{\mathcal{C}^* (r)}
 \liminf_{n \to \infty} \left( \frac{1}{n} \log L_n^* \right) \nonumber \\
 &=& \max_{q:D(q \parallel p) \leq r} H(q).
\end{eqnarray}
\end{cor}

\begin{figure}[ht]
 \begin{center}
  \includegraphics{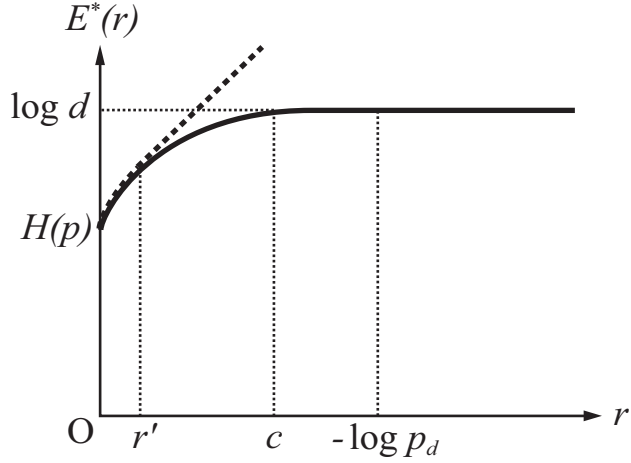}
 \end{center}
 \caption{\label{Fig4} Entanglement yield in entanglement concentration whose
  success probability exponentially decreases (strong converse).
  The horizontal axis represents the exponent of the success probability.
  The vertical axis represents the number of Bell pairs distilled per copy in
  the asymptotic limit:
  $E^*(r) = \max_{q:D(q \parallel p) \leq r} H(q)$.
  $E^*(r)$ reaches the maximum value, $\log d$, at $r=c= -\log d - d^{-1}\sum_i
  \log p_i$, which is the relative entropy between the uniform distribution
  $u \equiv (1/d, \cdots, 1/d)$ and $\{ p_i\}$.
  The broken line represents the entanglement yield $E^*_F(r)$ as a function
  of the exponent of the fidelity.
  At the value $r=r'$, where $\frac{d}{dr} E^* (r) = 1$, $E_F^* (r)$ starts to
  deviate from $E^*(r)$.}
\end{figure}

Entanglement yield can be related to fidelity also in the high-entanglement
regime, where the aim of the task is to convert $\ket{\phi}^{\otimes n}$
deterministically into some final (possibly mixed) state that is as close to a
maximally entangled state $\ket{\Phi_{T^*_n}}$ of size $T^*_n$ as possible.
We here require that the fidelity $F_n$ between the final state and the
maximally entangled state $\ket{\Phi_{T^*_n}}$ decrease as slowly as
$2^{-nr}$; namely,
\begin{equation} 
 r \geq \limsup_{n \to \infty} \left( -\frac{1}{n} \log F_n \right).
\end{equation}
Let us denote the class of all such sequences by $\mathcal{C}^*_F(r)$.
The maximum asymptotic entanglement yield $E^*_F(r)$ over $\mathcal{C}^*_F(r)$
can be obtained by reducing the problem to that of probabilistic schemes via
the following two lemmata.
\begin{lem}
 \label{lem:fidelity1}
 If the transformation $\ket{\phi} \longrightarrow \ket{\Phi_L}$ is possible
 with probability $\epsilon_1$, then, for any $T\geq L$, there exists a
 deterministic transformation $\ket{\phi} \longrightarrow \rho$ with 
 fidelity $\bra{\Phi_T}\rho\ket{\Phi_T}\geq \epsilon_1\epsilon_2$ with
 $\epsilon_2=L/T$.
\end{lem}
The proof is straightforward by considering the case
$\rho=\epsilon_1\ket{\Phi_L}\bra{\Phi_L}+(1-\epsilon_1) \rho'$,
by noting that $|\bra{\Phi_T} \Phi_L \rangle |^2= L/T$.

Applying this lemma to asymptotic sequences by setting
 $\epsilon_2=2^{-nx}$ where $x$ is an arbitrary nonnegative number,
we have the following: for any sequence of probabilistic schemes belonging to
$\mathcal{C}^*(r)$ with size $\{L_n^*\}$, there exists a sequence belonging to
$\mathcal{C}^*_F(r+x)$ with the size $\{T_n^*=L_n^* 2^{nx}\}$.
Hence, 
\begin{equation}
 \max_{\mathcal{C}^*_F(r+x)} \liminf_{n \to \infty} \left(\frac{1}{n}
 \log T_n^* \right)\geq E^*(r)+x, \;\; {}^\forall x \geq 0,
 \end{equation}
and therefore
\begin{equation}
 \max_{\mathcal{C}^*_F(r)} \liminf_{n \to \infty} \left(\frac{1}{n} \log T_n^*
 \right) \geq \sup_{0\leq x < r} \left\{ E^*(r-x)+x \right\}.
 \label{prob < fidelity:converse}
\end{equation}

\begin{lem}
 \label{lem:fidelity2}
 Suppose that there exists a deterministic transformation
 $\ket{\phi} \longrightarrow \rho$
 with fidelity $F=\bra{\Phi_T}\rho\ket{\Phi_T}$.
 Then, there exist an integer $L\leq T$ and a transformation
 $\ket{\phi} \longrightarrow \ket{\Phi_L}$ with success probability $P$,
 satisfying
 \begin{equation}
  \sqrt{PL} \geq \frac{\sqrt{TF}-1}{\ln(T)}.
 \end{equation}
\end{lem}
\begin{prf}
See \ref{App:Fidelity lemmata}.
\end{prf}
Using this lemma, we derive an upper bound of $\max_{\mathcal{C}_F^*(r)}
\limsup_{n \to \infty} \left(n^{-1} \log T_n^* \right)$.
Consider a sequence of deterministic schemes where the fidelity between the
final state of the $n$-th scheme and the maximally entangled state
$\ket{\Phi_{T_n^*}}$ is $F_n$.
Suppose that
\begin{equation} 
 r \geq \limsup_{n \to \infty} \left( -\frac{1}{n} \log F_n \right),
\label{koashilabel1}
\end{equation}
and
\begin{equation}
 \beta \equiv \limsup_{n \to \infty} \left(\frac{1}{n} \log T_n^* \right) > r.
 \label{koashilabel2}
\end{equation}

Let us take an infinite set of integers $\mathcal{N}$
such that 
$\lim_{n\rightarrow \infty; n \in \mathcal{N}} \left( n^{-1}
\log T_n^* \right) =\beta$.
According to Lemma~\ref{lem:fidelity2}, there exists a sequence of
transformations  $\ket{\phi}^{\otimes n} \longrightarrow
\ket{\Phi_{L_n^*}} (L_n^* \leq T_n^*)$ with success probability
$P^{(n)}_{\rm success}$ satisfying
\begin{equation}
 \sqrt{P_{\rm success}^{(n)} L_n^*} \geq \frac{\sqrt{T_n^* F_n}-1}
 {\ln(T_n^*)}.
\end{equation}
Since $\beta > r$, for any $\delta>0$, there exists $n_1$ such that for all
$n\geq n_1$,
\begin{equation}
 \frac{1}{n}(\log P^{(n)}_{\rm success} + \log L_n^*) \geq
 \frac{1}{n}(\log F_n + \log T_n^*)-\delta.
\end{equation}
Further, by equations~(\ref{koashilabel1}) and (\ref{koashilabel2}), there
exists 
$n_2\geq n_1$ such that for all $n\geq n_2$ with $n \in \mathcal{N}$,
\begin{equation}
 \frac{1}{n}(\log P^{(n)}_{\rm success} + \log L_n^*) \geq \beta-r-\delta.
\end{equation}
Let us define $r_1\equiv -\liminf_{n\rightarrow \infty; n \in \mathcal{N}}
\left( n^{-1}\log P^{(n)}_{\rm success} \right)$
(or, $r_1$ can be any accumulation value).
Note that
$L_n^* \leq T_n^*$ and the above inequalities imply
$r_1 \leq r$. Then, for any $\delta_1, \delta_2>0$, there
are infinite values of $n$ satisfying
$-\frac{1}{n}\log P^{(n)}_{\rm success} \leq r_1+\delta_1$ and
$n^{-1} \log L_n^* \geq \beta-r+r_1-\delta_2$.
This means that $E^*(r_1+\delta_1)\geq \beta+r_1-r-\delta_2$
holds for any $\delta_1, \delta_2>0$. Noting that
$E^*(r)$ is monotone increasing,
we have 
\begin{equation}
\lim_{x\downarrow r_1} E^*(x)\geq \beta-r+r_1.
\label{koashilabel3}
\end{equation}

Note that what we have shown here is that for any sequence
belonging to $\mathcal{C}_F^*(r)$ and having $\beta>r$, there exists a value
$r_1(\leq r)$
satisfying equation~(\ref{koashilabel3}). The same statement also holds for
sequences with $\beta \leq r$, since the left-hand side of
equation~(\ref{koashilabel3}) is nonnegative and the right-hand side is
nonpositive for $r_1=0$.
Hence,

\begin{equation}
 \max_{\mathcal{C}^*_F(r)} \limsup_{n \to \infty} \left(\frac{1}{n} \log T_n^*
 \right) \leq r+ \sup_{0\leq r_1 \leq r}  \left\{ \lim_{x\downarrow r_1} E^*(x)
 -r_1 \right\}.
\label{fidelity < prob:converse}
\end{equation}

Since $E^*(r)$ is continuous in $(0,\infty)$ and monotone increasing in
$[0,\infty)$, equations~(\ref{prob < fidelity:converse}) and
(\ref{fidelity < prob:converse}) determine
$E_F^*(r)$ as
\begin{eqnarray}
 E^*_F(r) &\equiv& \max_{\mathcal{C}^*_F(r)} \limsup_{n \to \infty} \left(
 \frac{1}{n} \log T_n^* \right) = \max_{\mathcal{C}^*_F(r)}
 \liminf_{n \to \infty} \left(\frac{1}{n} \log T_n^* \right) \nonumber \\
 &=& \sup_{0< x \leq r} \left\{ E^*(x)+r-x \right\}.
 \label{conversefidelityyield}
\end{eqnarray}
The function $E_F^*(r)$ is shown by the broken curve in figure~\ref{Fig4}.
Note that $\frac{\rm d}{{\rm d}r}E^*$ goes to $\infty$ at
$r\rightarrow 0$ and that it tends to $0$ as $r\rightarrow\infty$.
The value $r'$ in the figure satisfies $\frac{\rm d}{{\rm d}r}E^*=1$
and from this point $E_F^*(r)$ starts to deviate from $E^*(r)$ and
increases linearly.

In contrast to $E^*(r)$, $E_F^*(r)$ can reach any large value
by making $r$ large enough. 
This is because we are here allowing an exponentially small fidelity. Indeed,
for a separable initial state in which $E^*(r)=0$ everywhere, we still have
$E_F^*(r)=r$, which corresponds to the fact that separable states can attain
fidelity $2^{-N}$ to $N$ Bell pairs.
In the region of large $r$, the entanglement in the initial state
contributes to the vertical offset of $E_F^*(r)$,
which is given by $E_F^*(r')-r'$.

\section{\label{Sec:Alternative formulae}Alternative formulae}

Corollaries~\ref{cor:Direct cor} and \ref{cor:Converse2} give the entanglement
yields $E(r)$ and $E^*(r)$ as a minimum or a maximum over a probability
distribution $q$.
In practical calculation, however, it is not easy to deal with such
minimization or maximization over probability distributions.
We therefore present alternative formulae for entanglement yield, which contain
only minimization or maximization over a variable $s$.

The entanglement yields in Corollaries~\ref{cor:Direct cor} and
\ref{cor:Converse2} are also expressed as
\begin{equation}
 E(r) = \sup_{s\geq1} \frac{r + \psi(s)}{1-s},
 \label{alternative direct}
\end{equation}
and
\begin{equation}
 E^*(r) = \min_{0\leq s \leq1} \frac{rs + \psi(s)}{1-s},
 \label{alternative converse}
\end{equation}
where
\begin{equation}
 \psi(s) = \log \sum_i p_i^s.
\end{equation}
The proof of equations~(\ref{alternative direct}) and
(\ref{alternative converse}) is given in \ref{App:Another}.
The entanglement yields in the fidelity arguments can also be expressed in the
same manner:
\begin{equation}
 E_F(r) = \sup_{s\geq1} \frac{r + \psi(s)}{1-s},
 \label{alternative directfidelity}
\end{equation}
and
\begin{equation}
 E_F^*(r) = \left\{ \begin{array}{cc}
                   \min_{0\leq s \leq1} \frac{rs + \psi(s)}{1-s} &
                   \quad (0<r \leq r') \\
                   r - r' +E^*(r') & \quad (r'<r)
                   \end{array}
            \right.
 \label{alternative conversefidelity}
\end{equation}
where $r=r'$ is a value at which $\frac{d}{dr} E^*(r) = 1$.
Equations~(\ref{alternative directfidelity}) and
(\ref{alternative conversefidelity}) are easily obtained from
equations~(\ref{directfidelityyield}) and (\ref{conversefidelityyield}).

As an example derived  by using the above formulae, we show non-additivity of
the entanglement yield $E(r)$.
Additivity is an important property of
$E_{\rm entropy}(\rho)=H(p)$ and $E_{\rm det}(\rho)=-\log p_1$:
Entanglement of a composite system is the sum of the contributions from each
system, e.g.,
$E_{\rm entropy}(\rho \otimes \sigma)=E_{\rm entropy}(\rho)+
E_{\rm entropy}(\sigma)$.
(Note that $\rho$ and $\sigma$ represent pure states.)
This property, however, does not hold for the general expression of
entanglement yield, $E(r)$, as seen below.

We can prove the following inequality of $E_r$:
\begin{equation}
 E_{r_1 + r_2}(\rho \otimes \sigma) \leq E_{r_1}(\rho) + E_{r_2}(\sigma).
 \label{subadditivity}
\end{equation}
From equation~(\ref{alternative direct}), we have
\begin{eqnarray}
      E_{r_1 + r_2}(\rho \otimes \sigma)
  &=& \sup_{s\geq1} \frac{(r_1+r_2) + \psi_{\rho \otimes \sigma}(s)}{1-s}
      \nonumber \\
  &=& \sup_{s\geq1} \left\{ \frac{r_1 + \psi_{\rho}(s)}{1-s} +
      \frac{r_2 + \psi_{\rho}(s)}{1-s} \right\} \nonumber \\
  &\leq&   \sup_{s\geq1} \frac{r_1 + \psi_{\rho}(s)}{1-s} +
           \sup_{s\geq1} \frac{r_2 + \psi_{\sigma}(s)}{1-s} \nonumber \\
  &=& E_{r_1} (\rho) + E_{r_2} (\sigma).
\end{eqnarray}
Equality holds if and only if $s_+(r_1)_\rho = s_+ (r_2)_\sigma$, where
$s_+(r) \geq 1$ is a unique solution of $r= -\psi(s) -(1-s)\psi'(s)$
(see \ref{App:NewMonotonicity}).
Thus we obtain equation~(\ref{subadditivity}).

When $r_1=r_2=r/2$, equation~(\ref{subadditivity}) provides
\begin{equation}
 E_{r}(\rho \otimes \sigma) \leq E_{\frac{r}{2}}(\rho)+E_{\frac{r}{2}}(\sigma).
 \label{subadditivity2}
\end{equation}
Let the largest squared Schmidt coefficients of $\rho$ and $\sigma$ be $p_1$
and $q_1$, respectively.
Equality holds if and only if $\rho$ and $\sigma$ satisfy
$s_+(r/2)_\rho = s_+ (r/2)_\sigma$.
Note that this condition is satisfied when $r=0$ or $r \geq
-2\log ( \min\{p_1, q_1\} )$, which recovers the additivity of
$E_{\rm entropy}$ or $E_{\rm det}$, respectively.

Hence,
\begin{equation}
 E_{\frac{r}{2}}(\rho) = \frac{1}{2} E_r(\rho \otimes \rho).
 \label{half}
\end{equation}
Substituting this into equation~(\ref{subadditivity2}), we obtain
\begin{equation}
 E_r(\rho \otimes \sigma) \leq \frac{1}{2} \left\{ E_r (\rho \otimes \rho) +
  E_r(\sigma \otimes \sigma) \right\}.
 \label{average}
\end{equation}
This means that the entanglement yield in collective distillation of different
states does not exceed the average of those in separate distillations of
each state.
For $\rho$ and $\sigma$ such that
$s_+(r/2)_\rho \neq s_+ (r/2)_\sigma$,
the left-hand side of equation~(\ref{average}) is strictly less than the right
hand side.

On the other hand, consider the independent concentration of $\rho$ and
$\sigma$ with the same error exponent $r$.
The failure probability of the whole process is
\begin{equation}
 1-(1-2^{-nr})^2 = 2^{-nr+1} - 2^{-2nr}.
\end{equation}
Thus, the error exponent of the whole process is also $r$
(see \ref{App:greater exponent}).
Therefore, 
\begin{equation}
 E_r (\rho \otimes \sigma) \geq E_r (\rho) + E_r (\sigma),
\label{superadditive}
\end{equation}
which means that $E_r$ is generally non-additive.
In fact, equation~(\ref{half}) provides
\begin{equation}
 E_r(\rho \otimes \rho) = 2 E_{\frac{r}{2}} (\rho) > 2 E_r(\rho)
 \quad {\rm for} \quad r<-2\log p_1,
\end{equation}
thus, the inequality in equation~(\ref{superadditive}) is strict in this case.

To sum up, we obtain the following relation on the general entanglement yield
of composite pairs:
\begin{equation}
 E_r (\rho) + E_r (\sigma) \leq E_r (\rho \otimes \sigma)
 \leq \frac{1}{2} \left\{ E_r (\rho \otimes \rho)
  + E_r(\sigma \otimes \sigma) \right\}.
\end{equation}

\section{\label{Sec:Conclusion}Conclusion}

We have discussed entanglement concentration with exponentially decreasing
failure probability, as well as fidelity exponentially close to one, in the
asymptotic limit.
By employing the method of types, we derived the entanglement yield $E(r)$
as a function of an error exponent $r$.
The result fills the gap between the well-known least upper bound of
entanglement yield represented by entropy and the maximum attained in
deterministic concentration.
The explicit dependence on the exponent of the success probability and of
the fidelity was also presented, for the large-yield regime, in the form of a
strong converse.

In entanglement manipulation as well as other types of quantum information
processing, deterministic, probabilistic, and high-fidelity transformations
are considered.
Our results represent a common generalization and refinement of all three
approaches.
They provide a unified view of probabilistic and deterministic transformations,
and show that success probability and fidelity are essentially equivalent
concepts;
to be precise, high fidelity and high probability result in the same
yield function, while the yield function for low fidelity coincides
with that for low probability only for small exponents: for large
exponents the latter saturates whereas the former becomes a straight
line.
The power of the error rate approach (i.e., of quantifying `rare events')
in information theory is demonstrated also in quantum information theory.

\ack

MK was supported by a Grant-in-Aid for Encouragement of Young
Scientists (Grant No.~12740243) and a Grant-in-Aid for Scientific Research
(B) (Grant No.~12440111) by the Japan Society of the Promotion of Science.
KM is thankful to Professor Imai  for support.
FM is grateful to Takaaki Mukai for his support.
AW was supported by the U. K. Engineering and Physical Sciences Research
Council, and he acknowledges gratefully the hospitality of the ERATO Quantum
Computation and Information Project, Tokyo.

\appendix

\section{}
\label{App:greater exponent}

\begin{lem}
 \label{lem:limit}
 If \, $\lim_{n \to \infty} a_n \leq \lim_{n \to \infty} b_n$, \, then
 \begin{equation}
  \lim_{n \to \infty} -\frac{1}{n} \log (2^{-n a_n}+2^{-n b_n}) =
  \lim_{n \to \infty} a_n.
 \end{equation}
\end{lem}

\begin{prf}
 \begin{eqnarray}
 \fl \lim_{n \to \infty} -\frac{1}{n} \log (2^{-n a_n}+2^{-n b_n})
     &=& \lim_{n \to \infty} -\frac{1}{n} \log \left\{ 2^{-n a_n}
         \left( 1+2^{-n(b_n -a_b)} \right) \right\} \nonumber \\
     &=& \lim_{n \to \infty} \left\{ a_n - \frac{1}{n} \log \left (1
         +2^{-n(b_n -a_b)} \right) \right\} \nonumber \\
     &=& \lim_{n \to \infty} a_n.
 \end{eqnarray}
\end{prf}

\section{}
\label{App:NewMonotonicity}

\begin{lem}
 \label{lem:newmonotonicity}
Let $\mathcal{P}$ be a convex subset of $\mathbb{R}^n$.
Let $S: \mathcal{P} \to \mathbb{R}$ be a continuous function, and let
$T: \mathcal{P} \to \mathbb{R}$ be a strictly concave function.
Suppose that $T$ takes its maximum at $q_2 \in \mathcal{P}$ and
$S(q_2)=x_2$, and that
$\min_{q \in \mathcal{P}} S(q) = x_1$.
Then,
\begin{equation}
 R(x) \equiv \max_{q:S(q)=x} T(q) \qquad (x_1 \leq x \leq x_2)
\end{equation}
is strictly monotone increasing.
\end{lem}

\begin{prf}
Consider arbitrary $x'$ and $x''$ such that $x_1 \leq x' < x'' \leq x_2$.
Then, there exists $q'$ such that
$R(x') = \max_{q:S(q)=x'} T(q) = T(q')$ and $S(q') = x'$.
Due to the continuity of $S(q)$, there exists $q''$ such that $S(q'')=x''$.
Since the domain $\mathcal{P}$ is a convex set, we have
\begin{equation}
 q''=\lambda q_2 + (1-\lambda) q' \qquad (0 < \lambda \leq 1).
\end{equation}
Then, the strict concavity of $T(q)$ and our assumption provide
\begin{equation}
 T(q'') \geq \lambda T(q_2) + (1-\lambda) T(q') > T(q').
\end{equation}
Note that the strict concavity of $T$ ensures $T(q_2) > T(q')$.
Together with $T(q'') \leq \max_{q:S(q)=x''} T(q)$, we have
\begin{equation}
R(x') < R(x'') \qquad {\rm for} \qquad x' < x''.
\end{equation}
\end{prf}

We can easily derive the following corollary on \textit{convex} function.

\begin{cor}
 \label{cor:newmonotonicity2}
Let $\mathcal{P}$ be a convex subset of $\mathbb{R}^n$.
Let $S: \mathcal{P} \to \mathbb{R}$ be a continuous fuction, and let
$U: \mathcal{P} \to \mathbb{R}$ be a strictly convex function.
Suppose that $U$ takes its minimum at $q_2 \in \mathcal{P}$ and
$S(q_2)=x_2$, and that
$\min_{q \in \mathcal{P}} S(q) = x_1$.
Then,
\begin{equation}
 Q(x) \equiv \min_{q:S(q)=x} U(q) \qquad (x_1 \leq x \leq x_2)
\end{equation}
is strictly monotone decreasing.
\end{cor}

\section{}
\label{App:Fidelity lemmata}

This appendix gives the proofs of the lemmata used for reducing fidelity
arguments to probabilistic ones in Secs.~\ref{Sec:Asymptotic} and
\ref{Sec:Strong}.
Note that these lemmata are not confined to the analysis of
i.i.d.~entanglement concentration, but apply in the most general situations:
In fact, they apply whenever we are only interested in the \textit{rate} of the
entanglement produced and of the error probabilities (fidelities).

\begin{prf2}
\textbf{Proof of Lemma \ref{lem:lemma:1}} \
 According to Theorem 3 in reference~\cite{Vidal2000}, the optimum fidelity is
 always achieved with a deterministic transformation that actually produces a
 \textit{pure} state.
 So, our assumption implies that $\ket{\phi}$ can be deterministically
 transformed into a pure state $\ket{\phi'}$ such that
 $| \bra{\Phi_T}\phi' \rangle |^2 \geq 1- \epsilon$.
 We may assume (after suitable local rotations)
 \begin{eqnarray}
   \ket{\phi'}   &=& \sum_i \sqrt{p_i}\ket{i}\ket{i}, \\
   \ket{\Phi_T} &=& \sum_{i=1}^T \frac{1}{\sqrt{T}}\ket{i}\ket{i},
 \end{eqnarray}
 because this will only increase the fidelity.

 Then our assumption becomes
 \begin{equation}
  \sum_{i=1}^T \sqrt{p_i}\frac{1}{\sqrt{T}}
   \geq \left(\sum_{i=1}^T \sqrt{p_i}\frac{1}{\sqrt{T}}\right)^2
   \geq 1-\epsilon,
 \end{equation}
 from which we directly obtain
 \begin{equation}
  \sum_{i=1}^T \left(\sqrt{p_i}-\frac{1}{\sqrt{T}}\right)^2 \leq 2\epsilon.
 \end{equation}
 Suppose
 \begin{equation}
  p_1 \geq p_2 \geq \cdots \geq p_K \geq \frac{(1+\sqrt{2})^2}{T} > p_{K+1}
  \geq \cdots \geq p_T.
 \end{equation}
 (Note that if $p_1<(1+\sqrt{2})^2/T<6/T$, then we can
  deterministically obtain $\ket{\Phi_L}$ with $L=\left\lfloor
  T/6 \right\rfloor$ by the deterministic entanglement
  concentration \cite{Morikoshi2001} reviewed in section~\ref{Sec:Revisited}.)
 Then,
 \begin{equation}
 \fl \quad \frac{2 \epsilon}{T} \geq \frac{1}{T} \sum_{i=1}^T \left( \sqrt{p_i}
  - \frac{1}{\sqrt{T}} \right) ^2 \geq \frac{1}{T} \sum_{i=1}^K \left(
  \sqrt{p_i} - \frac{1}{\sqrt{T}} \right) ^2 \geq \frac{1}{T} \sum_{i=1}^{K}
  \frac{2}{T} = \frac{2K}{T^2}.
 \end{equation}
 Introducing $\beta \equiv K/T  \leq \epsilon$, and defining
 $\delta \equiv \sum_{i=1}^K p_i$, our aim is to bound the latter probability.
 
 Note that with the restriction that the largest $K$ probabilities $p_i$ sum
 to $\delta$, the fidelity to $\ket{\Phi_T}$ is maximized for
 \begin{eqnarray}
  \hat{p}_1=\ldots=\hat{p}_K     &=& \frac{\delta}{K}, \\
  \hat{p}_{K+1}=\ldots=\hat{p}_T &=& \frac{1-\delta}{T-K},
 \end{eqnarray}
 yielding
 \begin{eqnarray}
  1-\epsilon \leq F(\phi',\Phi_T) &\leq& F_{\rm max}(\delta,K) \nonumber \\
   &=& \left( K \sqrt{\frac{\delta}{K}} \frac{1}{\sqrt{T}}
        +(T-K) \sqrt{\frac{1-\delta}{T-K}} \frac{1}{\sqrt{T}} \right)^2 
        \nonumber \\
   &=& \left(\sqrt{\delta} \sqrt{\beta} +\sqrt{1-\delta} \sqrt{1-\beta}
       \right)^2 \nonumber \\
   &\leq& \sqrt{\delta} \sqrt{\beta}+\sqrt{1-\delta} \sqrt{1-\beta}.
 \end{eqnarray}
 If we denote the right-hand side by $1-\eta$, we can solve for
 $\sqrt{\delta}$, and obtain
 \begin{eqnarray}
  \sqrt{\delta} &=& \sqrt{\beta}(1-\eta) \pm \sqrt{(1-\beta)\eta(2-\eta)}
                     \nonumber \\
                &\leq& \sqrt{\beta} + \sqrt{2\epsilon} \nonumber \\
                &\leq& \bigl(1+ \sqrt{2} \bigr) \sqrt{\epsilon}.
 \end{eqnarray}
 Hence, $\delta\leq 6\epsilon$, and we can execute the following probabilistic
 protocol: First, Alice and Bob observe whether the state is in the subspace
 spanned by $\ket{1},\ldots,\ket{K}$ or in the orthogonal complement, using a
 projective measurement.
 The first result occurs only with probability $\delta\leq 6\epsilon$.
 In the other case, the post-measurement state has Schmidt coefficients
 $p_i'=p_i / (1-\delta)$.
 Hence,
 \begin{equation}
  \frac{6}{T(1-6\epsilon)}\geq p_{K+1}'\geq p_{K+2}'\geq\ldots ,
 \end{equation}
 so by the deterministic entanglement concentration \cite{Morikoshi2001},
 we can obtain $\ket{\Phi_L}$ deterministically,
 with $L=\left\lfloor T(1-6\epsilon)/6\right\rfloor$.
 Therefore, a pure state $\ket{\phi'}$, into which $\ket{\phi}$ is
 deterministically transformed, can be transformed into $\ket{\Phi_L}$ with
 probability $1-6 \epsilon$.
\end{prf2}

\begin{prf2}
\textbf{Proof of Lemma \ref{lem:fidelity2}} \ 
 According to Theorem 3 in reference~\cite{Vidal2000}, the optimum fidelity is
 always achieved with a deterministic transformation that actually produces a
 pure state.
 So, our assumption implies that $\ket{\phi}$ can be deterministically
 transformed into a pure state $\ket{\phi'}$ such that
 $| \bra{\Phi_T}\phi' \rangle |^2 = F$.
 Let $\ket{\phi'} = \sum_{i=1}^{M} \sqrt{p_i} \ket{i} \ket{i}$,
 and introduce the function $p(x)$ defined by 
 $p(x)=p_i \ (i-1 < x \leq i), \ p(x)=0 \ (M<x)$.
 Then, the fidelity $F$ to the state $|\Phi_T\rangle$ satisfies
  \begin{equation}
   \sqrt{TF}=\sqrt{p_1}+\int_1^T \sqrt{p(x)} dx
  \end{equation}
  Substituting $x=e^s$, we obtain
  \begin{equation}
  \sqrt{TF}-\sqrt{p_1}= \int_0^{\ln(T)} x\sqrt{p(x)} ds
                   \leq \ln(T) \max_{0\leq x\leq T}\left[x\sqrt{p(x)}\right].
  \end{equation}
  On the other hand, for any $x$, the probabilistic concentration
  with truncation value $p(x)$ gives a probabilistic
  transformation into $\ket{\Phi_L}$ with success
  probability $P$ with $L\geq x$ and $P\geq x p(x)$.
  (Note that $P=p(x)L$ due to equation~(\ref{revisitedfinal})
  in section~\ref{Sec:Revisited}.)
  Hence, $\sqrt{PL} \geq x \sqrt{p(x)}$.

  Combining those, we arrive at the conclusion that
  there exists a probabilistic scheme for some $L \leq T$
  satisfying
  \begin{equation}
  \sqrt{PL}\geq \frac{\sqrt{TF}-\sqrt{p_1}}{\ln(T)}
             \geq \frac{\sqrt{TF}-1}{\ln(T)}.
  \end{equation}
\end{prf2}

\section{}
\label{App:Another}

\begin{prf2}
\textbf{Proof of equations~(\ref{alternative direct}) and
(\ref{alternative converse})} \ 

 Consider the function
 \begin{equation}
  F(s) = -\psi(s) -(1-s) \psi' (s) \qquad (s \geq 0),
 \end{equation}
 where 
 \begin{equation}
  \psi(s) = \log \sum_{i=1}^d p_i^s,
 \end{equation}
 and $p_i$ is a probability distribution.
 Differentiating these functions, we have the following relations:
 \begin{equation}
  \psi' (s) = \sum_i h_i(s) \log p_i,
 \end{equation}
 where $h_i(s)$ is a probability distribution such that
 \begin{equation}
  h_i(s) = \frac{p_i^s}{\sum_j p_j^s},	
 \end{equation}
 \begin{equation}
  \psi''(s) = \ln 2 \left\{ \sum_i h_i (s) (\log p_i)^2 - \left(\sum_i h_i(s)
  \log p_i \right)^2 \right\} > 0,
  \label{2ndderivative}
 \end{equation}
 and
 \begin{equation}
  F'(s) = -(1-s) \psi''(s)
  \left\{
  \begin{array}{cc}
   >0 & \quad (s>1) \\
   =0 & \quad (s=1) \\
   <0 & \quad (0 \leq s<1).
  \end{array}
  \right.
 \end{equation}
 (Note that $\ln x$ denotes natural logarithm.)
 Since $F(1) =0$, $\lim_{s \to \infty}F(s)=-\log p_1$,
 and $F(0) = D(u \parallel p) \equiv c$, where $u=(1/d, \cdots, 1/d)$ is the
 uniform distribution, 
 there exist unique $s_+(r) > 1$ and $0 < s_-(r) < 1$
 such that
 \begin{equation}
  r = \left\{
      \begin{array}{cl}
       F(s_+ (r) ) & \quad r \in (0, -\log p_1) \\
       F(s_- (r) ) & \quad r \in (0, c).
      \end{array}
      \right.
  \label{Fvsr}
 \end{equation}

 Let $q$ and $h(s)$ be probability distributions such that $D(q \parallel p)=
 D(h(s) \parallel p) = r$.
 Then, we have
 \begin{eqnarray}
 \fl \quad H(h(s)) - H(q)
     &=& -D(h(s) \parallel p) - \sum_i h_i(s) \log p_i + D(q \parallel p)
         + \sum_i q_i \log p_i \nonumber \\
     &=& \sum_i \left\{ q_i - h_i(s) \right\} \log p_i,
 \end{eqnarray}
 and
 \begin{eqnarray}
 \fl \quad \frac{1}{1-s} D(q \parallel h(s))
  &=& \frac{1}{1-s} \left\{ D(q \parallel h(s)) + D(h(s) \parallel p)
      - D(q \parallel p) \right\} \nonumber \\
  &=& \frac{1}{1-s} \sum_i \left\{ -q_i \log h_i (s) +h_i (s) \log h_i (s)
      -h_i (s) \log p_i + q_i \log p_i \right\} \nonumber \\
  &=& \frac{1}{1-s} \left\{ (1-s) \sum_i (q_i - h_i (s)) \log p_i
      + \sum_i (q_i - h_i (s)) \log \sum_j p_j^s \right\} \nonumber \\
  &=& \sum_i \left\{ q_i - h_i (s) \right\} \log p_i.
 \end{eqnarray}
 Thus,
 \begin{equation}
  H(h(s)) - H(q) = \frac{1}{1-s} D(q \parallel h(s)),
  \label{HminusH}
 \end{equation}
 which gives
 \begin{equation}
  H(h(s_+(r))) \leq H(q) \qquad {\rm and} \qquad H(h(s_- (r)) \geq H(q).
  \label{minandmaxH}
 \end{equation}
 
 On the other hand,
 \begin{eqnarray}
  F(s) &=& -\log \sum_j p_j^s -(1-s) \sum_i h_i(s) \log p_i \nonumber \\
       &=& \sum_i h_i(s) \log h_i(s) - \sum_i h_i(s) \log p_i \nonumber \\
       &=& D(h(s) \parallel p),
 \end{eqnarray}
 hence,
 \begin{equation}
  D(h(s_{\pm}(r)) \parallel p) = F(s_{\pm}(r)) = r.
  \label{D=r}
 \end{equation}
 Therefore,
 \begin{eqnarray}
      \frac{s_{\pm} r + \psi(s_{\pm})}{1-s_{\pm}}
  &=& \frac{s_{\pm} F(s_{\pm}) + \psi(s_{\pm})}{1-s_{\pm}} \nonumber \\
  &=& \psi(s_{\pm}) -s_{\pm} \psi'(s_{\pm}) \nonumber \\
  &=& \log \sum_i p_i^{s_{\pm}} -s_{\pm} \sum_i h_i(s_{\pm}) \log p_i
      \nonumber \\
  &=& - \sum_i h_i(s_{\pm}) \log h_i(s_{\pm}) \nonumber \\
  &=& H(h(s_{\pm})).
  \label{H+-}
 \end{eqnarray}
 From equations~(\ref{minandmaxH}), (\ref{D=r}), and (\ref{H+-}), we obtain,
 for $r \in (0, -\log p_1)$,
 \begin{equation}
  \min_{q:D(q \parallel p) = r} H(q) = \frac{s_+ r + \psi(s_+)}{1-s_+},
  \label{preminDH}
 \end{equation}
 i.e.,
 \begin{equation}
  \min_{q:D(q \parallel p) = r} \left\{ D(q \parallel p) + H(q) \right\} =
  \frac{r + \psi(s_+)}{1-s_+},
  \label{minDH}
 \end{equation}
 and, for $r \in (0,c)$,
 \begin{equation}
  \max_{q:D(q \parallel p) = r} H(q) = \frac{s_- r + \psi(s_-)}{1-s_-}
  = \frac{r + \psi(s_-)}{1-s_-} - r.
  \label{maxH}
 \end{equation}
	 
Next, consider the function
\begin{equation}
 G(s) \equiv \frac{r + \psi(s)}{1-s}.
 \label{G}
\end{equation}
Then,
\begin{equation}
 G'(s) = \frac{r - F(s)}{(1-s)^2}.
\end{equation}
Since
\begin{equation}
 F(s) \left\{
      \begin{array}{cc}
       \geq r \quad & (s \leq s_-, \ s_+ \leq s) \\
       <r \quad & (s_- < s < s_+),
      \end{array}
      \right.
\end{equation}
we have
\begin{equation}
 G'(s) \left\{
       \begin{array}{cc}
        > 0 \quad & (s_- < s <s_+) \\
        = 0 \quad & (s=s_{\pm}) \\
        < 0 \quad & (s<s_-, \ s_+<s).
       \end{array}
       \right.
\end{equation}
(Note that we take $s_+ = \infty$ for $r \geq -\log p_1$.)
Thus, we obtain
\begin{equation}
 \max_{s\geq1} G(s) = G(s_+(r)) \quad {\rm for} \quad r \in (0, -\log p_1),
 \label{maxG}
\end{equation}
and
\begin{equation}
 \sup_{s\geq1} G(s) = \lim_{s \to \infty} G(s) = -\log p_1 \quad
  {\rm for} \quad r\geq -\log p_1.
  \label{supG}
\end{equation}
As discussed in section~\ref{Sec:Asymptotic}, the left-hand side of equation~
(\ref{minDH}) is monotone decreasing with respect to $r$.
Thus, together with equation~(\ref{G}), we obtain
\begin{equation}
\fl \min_{q:D(q \parallel p) \leq r} \left\{ D(q \parallel p) + H(q) \right\}
 = G(s_+) \quad {\rm for} \quad r \in (0, -\log p_1).
 \label{E=G}
\end{equation}
Therefore, equations~(\ref{maxG}), (\ref{supG}), and (\ref{E=G}) give 
equation~(\ref{alternative direct}).

On the other hand, for $r \in(0, c)$, we have $s_-(r) > 0$, thus
\begin{equation}
 \min_{0\leq s \leq1} G(s) = G(s_-(r)),
\end{equation}
hence,
\begin{equation}
 \min_{0 \leq s \leq 1} \frac{rs + \psi(s)}{1-s} =
 \min_{0 \leq s \leq 1} \left( G(s) -r \right) =
 G(s_-(r))-r.
\label{minG-r}
\end{equation}
In addition, for $r \geq c$, we have $s_-(r) \leq 0$, thus
\begin{equation}
 \min_{0\leq s \leq1} G(s) = G(0) = r + \log d,
\end{equation}
hence,
\begin{equation}
 \min_{0 \leq s \leq 1} \frac{rs + \psi(s)}{1-s} =
 \min_{0 \leq s \leq 1} \left( G(s) -r \right) = \log d.
 \label{G-r=logd}
\end{equation}
As discussed in section~\ref{Sec:Strong}, the left-hand side of equation~
(\ref{maxH}) is monotone increasing with respect to $r$.
Thus, together with equation~(\ref{G}), we obtain
\begin{equation}
 \max_{q:D(q \parallel p) \leq r} H(q) = G(s_-)-r \quad {\rm for} \quad
 r \in (0, c).
 \label{E*=G-r}
\end{equation}
Therefore, equations~(\ref{minG-r}), (\ref{G-r=logd}), and (\ref{E*=G-r}) give 
equation~(\ref{alternative converse}).

\end{prf2}


\section*{References}

\begin{thebibliography}{99}

\bibitem{Bennett92} Bennett C H and Wiesner S J 1992 \textit{\PRL} {\bf 69}
 2881

\bibitem{Bennett93} Bennett C H, Brassard G, Cr\'epeau C, Jozsa R, Peres A and
 Wootters W K 1993 \textit{\PRL} {\bf 70} 1895

\bibitem{Ekert} Ekert A K 1991 \textit{\PRL} {\bf 67} 661

\bibitem{Jozsa} Jozsa R and Linden N 2002 \textit{Preprint} quant-ph/0201143

\bibitem{Nielsen99} Nielsen M A 1999 \textit{\PRL} {\bf 83} 436, quant-ph/9811053

\bibitem{Vidal99} Vidal G 1999 \textit{\PRL} {\bf 83} 1046, quant-ph/9902033

\bibitem{Hardy99} Hardy L 1999 \textit{\PR}A {\bf 60} 1912, quant-ph/9903001

\bibitem{Jonathan99} Jonathan D and Plenio M B 1999 \textit{\PRL} {\bf 83} 1455,
 quant-ph/9903054

\bibitem{Bennett96} Bennett C H, Bernstein H J, Popescu S and Schumacher B
 1996 \textit{\PR}A {\bf 53} 2046, quant-ph/9511030

\bibitem{Popescu} Popescu S and Rohrlich D 1997 \textit{\PR}A {\bf 56} R3319,
 quant-ph/9610044

\bibitem{Lo} Lo H-K and Popescu S 2001 \textit{\PR}A {\bf 63} 022301,
 quant-ph/9707038

\bibitem{Morikoshi2000} Morikoshi F 2000 \textit{\PRL} {\bf 84} 3189,
 quant-ph/9911019

\bibitem{Morikoshi2001} Morikoshi F and Koashi M 2001 \textit{\PR}A {\bf 64}
 022316, quant-ph/0107120

\bibitem{Hayashi2001} Hayashi M and Matsumoto K 2001 \textit{Preprint}
 quant-ph/0109028

\bibitem{Hayashi2002} Hayashi M 2002 \textit{\PR}A {\bf 66} 032321,
 quant-ph/0202002

\bibitem{CT} Cover T M and Thomas J A 1991 \textit{Elements of Information
 Theory} (New York: John Wiley and Sons)

\bibitem{CK} Csisz{\'{a}}r I and K{\"{o}}rner J 1981 \textit{Information
 Theory: Coding Theorems for Discrete Memoryless Systems} (New York: Academic
  Press)

\bibitem{Vidal2000} Vidal G, Jonathan D and Nielsen M A 2000 \textit{\PR}A
 {\bf 62} 012304, quant-ph/9910099

\end{thebibliography}
\end{document}